%% file: Masterarbeit.tex
\documentclass[12pt]{article}
\usepackage{amsmath}    
\usepackage{graphics}   
\usepackage{graphicx} 
\usepackage[english]{babel}
\usepackage[latin1]{inputenc}
\usepackage{epsfig}
\usepackage[arrow, matrix, curve]{xy}
\usepackage{amsfonts}
\usepackage{amsthm}
\usepackage{color}
\usepackage{amssymb}
\usepackage{stmaryrd}
\usepackage{verbatim}

\begin {document}

\newcommand{\beq}{\begin{equation}}
\newcommand{\enq}{\end{equation}}

\newcommand{\HRule}{\rule{\linewidth}{0.5mm}}

\begin{titlepage}
\setcounter{page}{0}
\begin{center}
\HRule \\[0.4cm]
{ \huge \bfseries Variational Methods for Path Integral Scattering }\\[0.2cm]
\HRule \\[1.5cm] 
\begin{minipage}{0.4\textwidth}
\begin{flushleft} \large
\emph{Author:}\\
Julien \textsc{Carron}\\
\end{flushleft}
\end{minipage}
\begin{minipage}{0.4\textwidth}
\begin{flushright} \large
\emph{Supervisors:} \\
Roland \textsc{Rosenfelder}\\
 {\footnotesize PSI Villigen}\\
Jürg \textsc{Fröhlich}
\\{\footnotesize ETH Zürich}
\end{flushright}
\end{minipage}
\vspace{4cm}
\begin{abstract}
\noindent 
In this master thesis, a new approximation scheme to non-relativistic potential scattering is developed and discussed.
The starting points are two exact path integral representations of the T-matrix, which permit the application of the Feynman-Jensen variational method. A simple Ansatz for the trial action is made, and, in both cases, the variational procedure singles out a particular one-particle classical equation of motion, given in integral form. While the first is real, in the second representation this trajectory is complex and evolves according to an effective, time dependent potential. Using a cumulant expansion, the first correction to the variational approximation is also evaluated. The high energy behavior of the approximation is investigated, and is shown to contain exactly the leading and next-to-leading order of the eikonal expansion, and parts of higher terms. Our results are then numerically tested in two particular situations where others approximations turned out to be unsatisfactory. Substantial improvements are found.

\end{abstract}
\vfill
{\today}
\end{center}
\end{titlepage}

\tableofcontents

\newcommand{\vecx}{\mathbf x}
\newcommand{\vecxi}{\mathbf \xi}

\newcommand{\vecr}{\mathbf r}
\newcommand{\vecQ}{\mathbf Q}
\newcommand{\vecs}{\mathbf s}
\newcommand{\vecz}{\mathbf z}
\newcommand{\vecq}{\mathbf q}
\newcommand{\vecxt}{\mathbf x(t)}
\newcommand{\vecxs}{\mathbf x(s)}
\newcommand{\vecK}{\mathbf K}

\newcommand{\vecB}{\mathbf B}
\newcommand{\vecC}{\mathbf C}
\newcommand{\vecw}{\mathbf w}
\newcommand{\vecL}{\mathbf L}
\newcommand{\Kom}{\frac {\mathbf K} {m}}
\newcommand{\vecy}{\mathbf y}
\newcommand{\veck}{\mathbf k}

\newcommand{\vecv}{\mathbf v}
\newcommand{\vecb}{\mathbf b}

\newcommand{\vecp}{\mathbf p}

\newcommand{\gradt}{\nabla V(\vecxt)}
\newcommand{\grads}{\nabla V(\vecxs)}
\newcommand{\la}{\left\langle}
\newcommand{\rat}{\right\rangle}
\newcommand{\expd}{\left\langle\Delta S\rat}
\newcommand{\lp}{\left (}
\newcommand{\rp}{\right )}
\newcommand{\dpun}{\frac{d\vecp_1}{(2\pi)^3}}
\newcommand{\dpdeux}{\frac{d\vecp_2}{(2\pi)^3}}
\newcommand{\beqa}{\begin{eqnarray}}
\newcommand{\enqa}{\end{eqnarray}}
\newcommand{\Bv}{\left(\vecB,\vecv\right)}
\newcommand{\Cw}{\left(\vecC,\vecw\right)}
\newcommand{\intvw}{\int\mathcal D\vecv \mathcal D\vecw}
\newcommand{\sgn}{\mathrm{sgn}}

\newcommand{\xzero}{X_0}

\newcommand{\tautwo}{\chi_2^{(3-3)}}

\newcommand{\omtwovar}{\omega_2^{(3-3)}}

\newcommand{\dZZ}{\int\:dZ_{1,2}\:}
\newcommand{\dZZZ}{\int\:dZ_{1,2,3}\:}
\newcommand{\dt}[1]{\int\:dt_#1\:}
\newcommand{\dZ}[1]{\int\:dZ_#1\:}
\newcommand{\dx}[1]{\int\:d#1\:}
\newcommand{\pa}[1]{\partial_{#1}}
\newcommand{\V}[1]{V(r_{#1})}
\newcommand{\dabs}[2]{|Z_{#1}-Z_{#2}|}
\newcommand{\dsgn}[2]{\mathrm{sgn}(Z_{#1}-Z_{#2})}

\newcommand{\chir}[1]{\la\chi^{#1}_{ray}\rat}
\newcommand{\sumi}[3]{\sum_{#1=#2}^{#3}}
\newcommand{\prodi}[3]{\prod_{#1=#2}^{#3}}
\newcommand{\cumul}[1]{\lambda_{#1}}
\newcommand{\cumulr}[1]{\lambda_{#1}^{(3-1)}}
\newcommand{\omk}{\omega_{2}^{k_1}}
\newcommand{\omkk}{\omega_{2}^{k_2}}
\newcommand{\tauk}{\chi_{2}^{k_1}}
\newcommand{\taukk}{\chi_{2}^{k_2}}
\newcommand{\Vs}[1]{V_{\sigma(#1)}}

\newcommand{\C}{\mathrm C}

\newpage
\pagestyle{myheadings}
\markboth{Introduction}{J. Carron : Variational Methods for Path Integral Scattering}

\setcounter{equation}{0}
\section{Introduction}
The importance of variational methods in physics cannot be overestimated. The variational principle itself, which states that laws of nature are such that an appropriate function, or functional, is extremalized, can be safely traced back at least to ancient Greece. A that time, Hero of Alexandria formulated that the path of a light ray in a medium is the one that minimizes its length, and derived, from these considerations only, the reflection law in optics. In the 18th century, after works of Euler, Lagrange and others, the principle of least action was born. The principle states that the physical laws can be deduced from the condition of stationarity of a single quantity, called the action S,  
\beq
\delta S = 0.\nonumber
\enq
This approach proved so fruitful that nowadays, in most of modern physics, the starting point of almost any theory is an appropriate action, and the dynamical laws of the theory are determined through the requirement that the action is extremalized.\newline
\indent Unfortunately, very few physical problems can be solved exactly, but variational methods turn out to be very useful as approximation procedures, too. Certainly, there exists as many ways to approximate solutions to a given problem as problems themselves, but variational approximations are conceptually a class of their own, and can furnish fairly reliable results. Let us imagine that we are interested in a function $S[x]$, at a precise point $x_{\mathrm{ph}}$ that represents schematically the physical system of interest, but, for some reason, our knowledge of the system is incomplete, and we do not know what $x_{\mathrm {ph}}$ is. What we know is only that the quantity $S[x]$ is stationary at this particular point $x_{\mathrm {ph}}$. The very idea of this approximation is to restrict the range of values that can take $x$, namely, we require $x$ to lie on some particular subspace of the space in which $x$ took its value initially. Let us parametrize this subspace with the help of some parameters $a,b,$... so that $x \equiv x (a,b,...)$.
If the subspace in question is sufficiently broad, the point $x_t$ at which $S$ is stationary with respect to the parameters $a,b,$... may furnish a good approximation of $x_{\mathrm{ph}}$, and, what is more important, an even better approximation of $S[x_{\mathrm {ph}}]$, since this function is stationary a this point. We have thus reduced the task of finding $x_{\mathrm{ph}}$ to a simpler optimization problem. Nevertheless, success is not guaranteed, since it depends crucially on the chosen subspace, which may or may not be sufficiently close to the physical configuration.
\newline
\indent
Perhaps the most well-known variational approximation in quantum mechanics is the Ritz method to obtain approximate energy levels of bound quantum systems. An account can be found for example in \cite{Messiah}. To emulate the unknown wave function of a quantum system, a trial wave function $\Psi_t$ depending on some parameters is introduced. The true ground state energy $E_0$ of the system can be shown to be always lower than the expectation value of the system's Hamiltonian $\hat H$,
\beq
E_0 \leq \frac{\la \Psi_t|\hat H|\Psi_t\rat}{\la \Psi_t|\Psi_t\rat}.\nonumber
\enq
The optimization procedure then furnishes an approximative eigenstate, and a value for the corresponding energy that is even closer to the exact one. The fact that the true ground state energy always lies below the approximative value obtained, makes this method especially useful. More than a variational principle, is is a minimum principle. 
\newline\newline 
\indent
In this master thesis, we apply such a variational method in the framework of potential scattering, in non-relativistic quantum mechanics. This is an old, time-honored subject whose treatment can be found in many textbooks \cite{Newton},\cite{Rodberg}. Potential scattering theory considers the elastic scattering of a quantum mechanical particle, that interacts through a local potential, which itself falls off sufficiently rapidly, so as to permit the existence of free initial (long before the scattering) and final (long after the scattering) states. The objects of interest are the scattering matrix (S-matrix), that expresses the probabilities of a transition between given initial and final states, and (or) the closely related T-matrix. Potential scattering theory has the advantage of being numerically solvable through the partial-wave expansion. We use therefore this framework in this thesis as a test case for more complicated processes, such as many-body scattering. \newline\indent
Approximation methods in potential scattering theory are numerous. To mention some, a truncated partial-wave series is very useful for low energies, while the eikonal expansion\cite{Wallace} or the Born expansion cover the high-energy range. 
It has to be said that variational methods in potential scattering exist already, too. One could cite for example a method introduced by Kohn \cite{Kohn} in 1948, that is now abundantly used in electron or positron scattering from atoms or molecules. Also, a stationary expression for the T-matrix was given by Schwinger \cite{Schwinger}, which can be used to calculate the partial-wave phase shifts. So, what brings this thesis that is new?
\newline\newline\indent
The starting point are two exact path-integral representations of the T-matrix, that were very recently published \cite{Rosenfelder}. The path-integral formulation of quantum mechanics was first worked out in the 1940's by Richard Feynman \cite{Feynman}, based upon a paper by Paul Dirac on the role of the Lagrangian in quantum mechanics. This formulation brought new insights into the structure of quantum physics, and proved essential to the subsequent development of theoretical physics. In this formulation, the probability of an event is explicitly written as a sum over all paths, that is over each and every possibility that leads to that event. Each path carries a probability amplitude, a complex number, that has the same magnitude for each path, but has a different phase, given by $i$ times the ratio of the action $S$ to the reduced Planck's constant $\hbar$. This formulation naturally contains classical mechanics, in the sense that if the action is much larger than $\hbar$, the paths that contribute most to the realization of the event are those for which the action is stationary, i.e. the classical paths. Besides, many applications were subsequently found in statistical physics or quantum field theories, where it is now a major tool \cite{RosenfelderII}\cite{Mosel}. Path integrals are convenient in the sense that they deal with classical quantities and usual numbers instead of quantum mechanical operators. A variational principle for actions in a path integral formalism is provided by the Jensen inequality for convex functions. With the help of a trial action $S_t$, one shows that
\beq
\la e^{-\Delta S/\hbar}\rat \geq e^{-\la \Delta S/\hbar\rat},\nonumber
\enq 
where the average is taken with respect to the positive, convex, weight function $e^{-S_t/\hbar}$, and $\Delta S$ stands for the difference $S-S_t$. This leads, as in the case of the Ritz method, to a minimum principle, that is used for instance in statistical mechanics, and was first introduced by Feynman. The main restriction to the method is the fact the trial action must be simple enough to allow analytical computations to be pursued. Corrections to this approximation can also be evaluated.
\newline\indent
The two path-integral representations mentioned above for the T-matrix are free of any seemingly diverging phases in the limit of asymptotic times, and instead avoid the need of infinite limits that occurred in previous descriptions, which rendered identities rather formal and not well suited for practical calculations. While it was shown in \cite{Rosenfelder} that these representations lead immediately to a new high-energy expansion scheme, they have the virtue that a variational procedure is also possible. This particular variational approximation is very similar to the one we just described above, and has the valuable advantage over the previous variational approximations we quoted, that it is not tied to the partial wave formalism, nor is restricted to high energies only. Also, it is the action that is forced upon the optimization procedure, and not a wave function, or a T-matrix element; it forms thus a completely new way to address the scattering problem, which permits to gain new insights. The difference with the Feynman-Jensen procedure is, as we will explain in more a detailed fashion later, that we will deal with complex quantities, i.e scattering amplitudes. As a consequence, our variational principle will not be a minimum principle anymore. This is a property also shared by the other quantum mechanical variational principles of Kohn and Schwinger. 
\newline\newline\newline
\noindent
The structure of the thesis is the following. In section 2, the stage is set up: we derive the two exact path-integral representations of the T-matrix. The derivation follows closely \cite{Rosenfelder}. In section 3, a stationary expression for the T-matrix is written down, the Feynman-Jensen variational principle and our variational Ansatz are presented in some more details. Corrections are also discussed. In section 4 and 5, the variational approximations to the T-matrix corresponding to the two representations are performed and discussed. The first correction to the approximation is also given. Numerical tests for two potentials are then presented in section 6, followed by our conclusions and outlook. Some technical computations are left for the appendices. 
\newline\newline
In this work, we use natural units in which $\hbar = 1$.
 
\setcounter{equation}{0}
\newpage
\input{PathTMatrix.tex}

\setcounter{equation}{0}
\input{F-JPrinciple}

\setcounter{equation}{0}
\input{EikonalRepresentation.tex}

\setcounter{equation}{0}
\input{RayRepresentation.tex}

\setcounter{equation}{0}
\input{Numerics.tex}

\setcounter{equation}{0}
\newpage
\section{Summary and Outlook}
In this master thesis, we have investigated a new variational approach to potential scattering, starting from two path-integral representations of the T-matrix, where one integrates functionally over velocities. Although it shares some common properties with the eikonal approximation, it possesses new and interesting features, and has a greater domain of validity. The Feynman-Jensen variational principle we used, together with our choice of trial action (the free action plus a linear term to be determined), select the best reference trajectory the particle travels along, given in the form of an integral equation of motion. The T-matrix is then given by summing up phases along this path, as in the eikonal approximation. Unlike its eikonal counterpart, this trajectory is in the first representation a real scattering trajectory that obeys Newton's equation of motion. In the second representation, interestingly, the path is complex, and obeys a similar classical equation of motion, albeit under the influence of a new, effective potential which has an explicit dependence on time. The high-energy behavior of the approximations was then investigated. Although the recovery of the leading and next-to-leading terms of the eikonal expansion was expected, the fact that the second order imaginary term was also found came as a pleasant surprise. Numerical tests of this approximation in a high energy situation have then shown that it does better at large angles than the eikonal approximation of Abarbanel and Itzykson. The numerical evaluation of the first correction to the variational result (the second cumulant) turned out to be very challenging. However, it became feasible with the use of adaptive integration procedures. This was really worth a try, since, especially in the second representation, the agreement between the exact results and our approach is remarkable.
\newline\newline\noindent
Naturally, this method also has its shortcomings. For example, the choice of trial action is restricted to those at most quadratic in the velocities. We studied in this work one simple choice of linear Ansatz, with the physical motivation that such a linear term in the T-matrix would emulate the structure of the high energy expansion, albeit without being subjected to high energy conditions. It would be interesting to study another, more general kind of quadratic trial action, where one also determines the quadratic part of the trial action from the variational principle. Inevitably, this will lead to more involved, coupled equations, that are more difficult to solve. Also, we have been unable to find a way to analyse the behavior of the approximation for low energies, although one may expect that the variational approximation could also be meaningful in this kinematic regime.\newline
As further investigations of this formalism one could propose the extension to relativistic scattering, as well as applications in many-body processes.

\addcontentsline{toc}{section}{Appendices} 

\newpage
\setcounter{equation}{0}

\input{Appendices.tex}

\newpage

\bibliographystyle{unsrt}
\bibliography{Masterarbeit}
\addcontentsline{toc}{section}{References} 
\end{document}

%% file: PathTMatrix.tex
\section{Path Integrals for the T-Matrix}\label{pathint}
In the framework of nonrelativistic potential scattering, we consider a potential $V(\mathbf r)$, that vanishes at infinity, initial and final free states $\phi_i$ and $\phi_f$, to which is associated an initial and final momentum $\veck_i$ and $\veck_f$. The S-matrix is the matrix element, taken between these free states, of the quantum mechanical propagator in the interaction picture, evaluated at asymptotic times,
\beq
S_{i\rightarrow f} = \lim_{T\rightarrow\infty}\la\phi_f\left|\hat U_I(T,-T)\right|\phi_i\rat.
\enq
The free states are normalized through $\la\phi_f|\phi_i\rat = (2\pi)^3\delta(\veck_f-\veck_f)$. The definition of the propagator is
\beq
\hat U (t_b,t_a) = \exp \lp -i \hat H(t_b-t_a)\rp,
\enq 
while in the interaction picture it reads
\beq
\hat U_I (t_b,t_a) = \exp\lp i\hat H_0t_b\rp\exp \lp -i \hat H(t_b-t_a)\rp  \exp\lp i\hat H_0 t_a \rp.
\enq 
From the S-matrix one usually substracts the identity, that does not represent actually the scattering process, and factors out an energy conserving Dirac delta function. What is left is called the T-matrix:
\beq
S_{i\rightarrow f} := (2\pi)^3\delta(\veck_i-\veck_f) -2\pi\:i\delta(E_i-E_f)T_{i\rightarrow f}.
\enq
We can now start the derivation of the path integral formulation of the T-matrix. We will follow closely \cite{Rosenfelder}, and present the main points of the derivation. We refer to it for a more complete discussion. The starting point is the Feynman-Kac formula, that presents the matrix elements
\beq U(t_b,\vecx_b,t_a,\vecx_a) := \la\vecx_b\left|\exp\lp -i\hat H(t_b-t_a)\rp\right|\vecx_a\rat\enq of the propagator between eigenstates of the $\vecx$ operator as a functional integration over paths. It formally reads
\beqa
U(t_b,\vecx_b,t_a,\vecx_a) &=& \lim_{N\rightarrow\infty}\lp \frac{m}{2\pi i\epsilon}\rp^{\frac{3N}{2}}\int d\vecx_1\cdots d\vecx_{N-1} \\
&\cdot&\exp\lp i\epsilon\sumi j0{N-1} \left\{ \frac{m}{2}\left[\frac{\vecx_{j+1}-\vecx_j}{\epsilon^2}\right]^2- V(\vecx_j)\right\} \rp.
\enqa
In this expression $\vecx_0$ is meant to be $\vecx_a$, and $\vecx_N$ is $\vecx_b$. This identity relies on the subdivision of the time intervall $(t_b-t_a)$ into $N$ small steps of duration $\epsilon = (t_b-t_a)/N$, and the insertion of a complete set of $\vecx$ eigenstates between each such step. One thus integrates over all possible intermediate values between $\vecx_a$ and $\vecx_b$. We will however use a different version of this formula, i.e. one that uses integration over velocities, and not paths.
\paragraph{Velocity Path Integrals}
This is done by inserting a "big one" in the $\vecx_i$ integrations, in the form of
\beq
1 = \prodi k1N\int d\vecv_k\:\delta\left(\vecv_k - \frac{\vecx_k -\vecx_{k-1}}{\epsilon}\right).
\enq 
The $\delta$-functions so introduced enforce sucessively
\beq
\vecx_j  = \vecx_0 + \epsilon\sumi i1j\vecv_j,
\enq
and permit to effectuate each $\vecx_i$ integration. However, from the $N$ $\delta$-functions, one is still left, that enforces
\beq
\vecx_b = \vecx_a + \epsilon \sumi i1N\vecv_i ,\quad \mathrm {or} \quad \vecx_b = \vecx_a + \int_{t_a}^{t_b} dt\:\vecv(t),
\enq
in continous notation. One has therefore
\beqa
U(t_b,\vecx_b,t_a,\vecx_a) &=&\nonumber \lim_{N\rightarrow\infty}\lp \frac{m}{2\pi i\epsilon}\rp^{\frac{3N}{2}}\int d\vecv_1\cdots d\vecv_{N}\:\delta\lp \vecx_b-\vecx_a-\epsilon\sumi j1N\vecv_j \rp \\
&\cdot&\exp\lp i\epsilon\sumi j1{N} \left\{ \frac{m}{2}\vecv_j^2- V\lp\vecx_a + \epsilon\sumi i1j\vecv_i \rp\right\} \rp.
\enqa
From now on we will exclusively employ the continous version of formulae of this type, that is, in this case,
\beqa 
U(t_b,\vecx_b,t_a,\vecx_a) \nonumber&=& \int \mathcal D\vecv \:\delta\lp \vecx_b-\vecx_a-\int_{t_a}^{t_b}dt\:\vecv(t)\rp \\
&\cdot&\exp\lp i\int_{t_a}^{t_b} dt\:\left\{ \frac{m}{2}\vecv^2(t)- V\lp\vecx(t)\rp\right\} \rp.\label{velocity}
\enqa
Here the $\mathcal D \vecv$ term is defined in the discrete version as
\beq
\mathcal D\vecv =\lp \frac{\epsilon m}{2\pi i} \rp ^{\frac{3N}{2}} \prodi k1Nd\vecv_k. 
\enq
It has the property that the pure gaussian functional integral is normalized to unity. The argument $\vecx(t)$ of the potential is 
\beq
\vecx(t) = \vecx_a + \int_{t_a}^{t} dt\:\vecv(t) = \frac{\vecx_a+\vecx_b}{2} + \frac12 \int_{t_a}^{t_b}ds\:\sgn(t-s)\vecv(s).
\enq
This identity can be inserted into the S-matrix. Indeed, after recognizing that
\beq
S_{i\rightarrow f} = \lim_{T\rightarrow\infty}e^{i\lp E_i+E_f \rp T}\la\phi_f\left| \hat U(T,-T) \right|\phi_i  \rat,
\enq
one inserts two complete sets of eigenstates $\vecx_a$ and $\vecx_b$ of the $\vecx$ operator, and recovers the matrix element $U(t_b,\vecx_b,t_a,\vecx_a)$. Together with (\ref{velocity}), it results in
\beqa
&&S_{i\rightarrow f} = \lim_{T\rightarrow\infty}e^{i\lp E_i+E_f \rp T}\int d\vecx_a\int d\vecx_b\:\exp\lp -i\left[\veck_f\cdot\vecx_b + \veck_i\cdot\vecx_a \right] \rp\\
&\cdot& \int \mathcal D\vecv \:\delta\lp \vecx_b-\vecx_a-\int_{-T}^{+T}dt\:\vecv(t)\rp 
\exp\lp i\int_{-T}^{+T} dt\:\left\{ \frac{m}{2}\vecv^2(t)- V\lp\vecx(t)\rp\right\} \rp\nonumber.
\enqa
After the change of variables $\vecr = (\vecx + \vecy)/2$ and $\vecs = \vecx -\vecy$, the relative coordinate $\vecs$ is fixed by the $\delta$-function. We furthermore define the mean momentum and momentum transfer through 
\beq
\vecK = \frac12\lp\veck_i+\veck_f\rp, \quad\quad \vecq = \veck_f-\veck_i,\\
\enq
and obtain the representation
\beqa
S_{i\rightarrow f} &=& \lim_{T\rightarrow\infty}e^{i\lp E_i+E_f \rp T}\int d\vecr\:e^{-i\vecq\cdot\vecr}
\int \mathcal D\vecv \:
\exp\lp i\int_{-T}^{+T} dt\:\frac{m}{2}\vecv^2(t) \rp\nonumber\\
&&\cdot \exp\lp-i \int_{-T}^{+T} \vecK\cdot\vecv(t)+ V\lp\vecr + \vecx_v(t)\rp\rp,
\enqa
where we used
\beq
\vecx_v(t) = \frac12 \int_{-T}^{+T}ds\:\sgn(t-s)\vecv(s).\enq
With a simple shift one gets rid of the linear term $\vecK\cdot\vecv(t)$. Indeed, with $\vecv(t) \rightarrow \vecv(t) +\vecK/m$, recognizing that $E_i+E_f - \vecK^2/m = \vecq^2/4m$, and that
\beq
\int_{-T}^{+T}ds\:\sgn(t-s) = 2t,\quad t\in [-T,T],
\enq
one has finally
\beqa
S_{i\rightarrow f} &=& \lim_{T\rightarrow\infty}\exp{\left(i\frac{\vecq^2}{4m}T\right)}\int d\vecr\:e^{-i\vecq\cdot\vecr}
\int \mathcal D\vecv \:
\exp\lp i\int_{-T}^{+T} dt\:\frac{m}{2}\vecv^2(t) \rp\nonumber\\
&&\cdot \exp\lp-i \int_{-T}^{+T} V\lp\vecr +  \frac\vecK m t+\vecx_v(t)\rp\rp.
\enqa
For a vanishing potential one calculates from this expression
\beq
S_{i\rightarrow f} = (2\pi)^3 \delta(\veck_i-\veck_f),
\enq
so the part of the S-matrix that describes the scattering is
\beqa
(S-1)_{i\rightarrow f} :=&& \lim_{T\rightarrow\infty}\exp{\left(i\frac{\vecq^2}{4m}T\right)}\int d\vecr\:e^{-i\vecq\cdot\vecr}
\int \mathcal D\vecv \:
\exp\lp i\int_{-T}^{+T} dt\:\frac{m}{2}\vecv^2(t) \rp\nonumber\\
&&\cdot \left\{\exp\lp-i \int_{-T}^{+T} V\lp\vecr + \frac\vecK m t +\vecx_v(t)\rp\rp-1\right\}.
\enqa
\paragraph{Diverging Phase}
Is it possible to get rid of the ill-defined phase $T\vecq^2/(4m)$ in front of the scattering matrix? To this aim, the factor $\vecq^2$ is transformed into a 3-dimensional Laplacian $\Delta$ acting on the variable $\vecr$, in the exponent of $\exp \lp-i\vecq\cdot\vecr \rp$. Since we are assuming that the potential vanishes at infinity, we can then integrate by parts without any boundary terms and let it act on the potential term. One can further reduce it to a shift operator with the help of
\beq
\exp\lp-\frac{i}{4m}T\Delta\rp = \int \mathcal D \vecw \exp\left( -i\int_{-T}^{+T}dt\:\frac{m}{2}\vecw^2(t)\pm\int_{-T}^{+T}dt\:\frac12f(t)\vecw(t)\cdot\nabla\right).
\enq
In this expression, $f(t)$ can be any function, as long as it fulfills the condition
\beq
\int_{-T}^{+T} dt\:f(t) = 2T,
\enq
so as to recover $T$ times the Laplacian, and the functional integration is normalised as before, that is, the pure gaussian integral is the identity.
The additional minus sign in the quadratic term is mandatory in order to get a real shift operator acting on the potential, while the linear term can have any sign. For reasons of convenience we choose the minus sign, and $f(t)$ to be $\sgn(-t)$. In that case the S-matrix becomes
 
\beqa
(S-1)_{i\rightarrow f} :=&&\lim_{T\rightarrow\infty}\int d\vecr\:e^{-i\vecq\cdot\vecr}\nonumber
\int \mathcal D\vecv\int\mathcal D\vecw \:\exp{\lp i\int_{-T}^{+T} dt\:\frac{m}{2}\left[\vecv^2(t)-\vecw^2(t)\right] \rp}\\ 
&&\cdot \left\{\exp\lp-i \int_{-T}^{+T} V\lp\vecr + \frac \vecK m t+ \vecx_v(t) -\vecx_w(0)\rp\rp-1\right\}. \label{path}
\enqa
These new degrees of freedom, that appear with a kinetic term of the wrong sign, thus remove all infinities. The 3-dimensional vector $\vecw$ will be called antivelocity on some occasions.
\paragraph{Extraction of the T-Matrix}
In order to reach for the T-matrix, one needs to extract the energy conserving $\delta$-function from (\ref{path}). This is done first by observing that the action in the path integral is (at least formally) invariant under the transformation
\beq \label{gauge}
t = \bar{t} + \tau, \quad \vecr = \bar \vecr -\frac{\vecK}{m}\tau,\quad \mathrm {and}\quad \vecv(t) = \bar{\vecv}(\bar t). 
\enq
Indeed, such a transformation only affects the integration bounds,
\beqa \nonumber
&&\int_{-T}^{+T}dt\: V\lp \vecr+\frac \vecK m t +\vecx_v(t)-\vecx_w(0)\rp \\
&=& \int _{-T-\tau}^{+T -\tau}d\bar t \:V\lp\bar{\vecr} + \frac{\vecK}{m}\bar t +\frac 12 \int _{-T-\tau}^{+T-\tau}d\bar s\:\bar{\vecv}(\bar s)\sgn(\bar t-\bar s)-\vecx_w(0) \rp.
\enqa
We expect that in the infinite T limit, this change of the bounds does not modify the action. This also means that it does not depend on the component of $\vecr$ that is parallel to $\vecK$. The integration over it thus produces a singularity, which is the energy conserving $\delta$-function we are looking for. To extract it, we use the Faddev-Popov trick, i.e. we insert first the "gauge fixing" term
\beq
1 = \frac{K}{m}\int d\tau\:\delta\lp\hat\vecK\cdot\left[\vecr + \frac \vecK m\tau\right]\rp,
\enq
in the $\vecr$ integration of (\ref{path}). The vector $\hat \vecK = \vecK/K$ is the unit vector in the $\vecK$ direction.
We then perform the transformation (\ref{gauge}), which, as we argued, does not change the action, but only the $\exp (-i\vecq\cdot\vecr)$ in the $\vecr$ integration. The only dependence on $\tau$ is now to be found in
\beqa \nonumber
&&\int d\tau \int d\vecr\:\exp\lp -i\vecq\cdot\vecr +i\:\vecq\cdot\frac\vecK m\tau\rp\delta\lp\hat\vecK\cdot\vecr\rp\\
&=& (2\pi)\delta\lp\frac{\vecq\cdot\vecK}{m}\rp\int d^2b\:\exp\lp-i\vecq\cdot\vecb\rp
\enqa 
As desired, the argument of the $\delta$-function is the difference in energy before and after the scattering. The 2-dimensional vector $\vecb$ (the impact parameter) is the part of $\vecr$ that is perpendicular to $\vecK$. 
As a consequence of the conservation of energy, the relations
\beq
K = 2k\cos\left(\frac\theta 2\right),\quad q = k\sin\left(\frac\theta 2\right)
\enq
hold, where $\theta$ is the scattering angle.
\paragraph{Eikonal Representation of the T-Matrix}
We have thus found our first representation of the T-matrix. It now reads
\beqa 
T^{(3-3)}_{i\rightarrow f}\nonumber &=& i\frac K m\int d^2b\:e^{-i\vecq\cdot\vecb}\intvw\:\exp\lp i \frac m2 \int dt\: \left[\vecv^2(t)-\vecw(t)^2\right]\rp \nonumber\\ 
&&\cdot \lp\exp\left[-i\int dt\:V(\vecxi_\vecK(t)) \right]-1\rp,\label{T33}
\enqa
where the trajectory $\mathbf\vecxi_\vecK$ is 
\beq 
\vecxi_\vecK(t) := \vecb + \frac\vecK m t +\vecx_v(t) -\vecx_w(0).
\enq
The upperscript (3-3) indicates the presence of the two functional integrations over the 3-dimensional variables $\vecv$ and $\vecw$. We will see that it is possible to write a similar formula with an 1-dimensional antivelocity $\vecw$ only.\newline The structure of the T-matrix is transparent: the reference trajectory along which phases are collected is the straight line, eikonal trajectory $\vecb + t\:\vecK/m$, while the functional integrals describe the quantum fluctuations around this path. We will use many times the notation
\beq
\chi_\vecK(\vecb,\vecv,\vecw) := -\int dt\:V\lp\mathbf\vecxi_\vecK(t)\rp.
\enq  
\paragraph{Ray Representation of the T-Matrix}
The full 3-dimensional path integral over $\vecw$ was introduced with the hope to remove all infinities. We will see in the subsequent chapters that this works in a very efficient way. However, it is also possible to use a one-dimensional, longitudinal, functional integration only, which has the effect of changing the reference trajectory. Let us consider in (\ref{T33}) the following shifts of variables:
\beqa
\vecv(t) &=& \frac{\vecq}{2m}\sgn(t)+\hat\vecv(t),\quad\quad \vecw(t) = \frac\vecq{2m}\sgn(t)+\hat\vecw(t),\\
\mathrm{and}\quad \vecb &=& \hat\vecb -\vecx_{\hat v\perp}(0)+\vecx_{\hat w\perp}(0).\label{bshift} 
\enqa
A small calculation shows that, on one hand,
\beq\frac m2
\int dt\:\left[\vecv^2(t)-\vecw^2(t)\right] = \frac m2\int dt\:\left[\hat\vecv^2(t)-\hat\vecw^2(t)\right] + \vecq\cdot\left[\vecx_{\hat w\perp}(0)-\vecx_{\hat v\perp}(0)\right].
\enq
This is to say that the shift (\ref{bshift} ) of the impact parameter exactly compensates for the difference and we recover the same free action. On the other hand, using the relation
\beq \label{sgn}
\int_{-T}^{+T}ds\:\sgn(t-s)\sgn(u-s) = 2T -2|t-u|, \quad t,u \in [-T,T],
\enq
the argument of the potential, that we will call now $\mathbf\vecxi_{ray}(t)$, becomes
\beq
\vecxi_{ray}(t) = \hat\vecb +\frac\vecK m t + \frac{\vecq}{2m}|t| + \vecx_{\hat v\perp} (t)-\vecx_{\hat v\perp}(0)-\vecx_{\hat w\sslash}(0).
\enq
The dependence on the components of $\vecw$ that are perpendicular to the mean momentum $\vecK$ is thus to be found only in the free part of the action. Given our normalisation of $\mathcal D \vecw$, we can simply forget about them.
The new reference path is now
\beq
\vecb + \frac\vecK m t+ \frac\vecq{2m}|t|\: := \:\vecb + \frac{\vecp_{ray}(t)}m t \: = \vecb +  \:\veck_i\Theta(-t) + \veck_f\Theta(t).
\enq 
A particle on this trajectory follows a straight line, according to its initial free motion, suffers, at $\vecb$ and at time $t = 0$, an elastic bump and moves again freely, in the direction given by its final momentum $\veck_f$. This path may intuitively represent better the picture of the scattering process than the straight-line eikonal path we have just seen. Besides, it has the advantage that the particle has the right scattering energy $k^2/(2m)$ instead of the slightly misplaced $K^2/(2m)$ in the eikonal version.
We call this representation of the T-matrix the ray representation. It reads
\beqa 
T^{(3-1)}_{i\rightarrow f}\nonumber &=& i\frac K m\int d^2b\:e^{-i\vecq\cdot\vecb}\int \mathcal D\vecv\int \mathcal D w\:\exp\lp i \frac m2 \int dt\: \left[\vecv^2(t)-w(t)^2\right]\rp\\ \label{T31}
&&\cdot \left\{\exp\left [i\chi_{ray}\left(\vecb,\vecv,w\right)\right]-1\right\},
\enqa
with the phase
\beq \chi_{ray} (\vecb,\vecv,w) = -\int dt\:V\lp\vecxi_{ray}(t)\rp.\enq
\noindent
\paragraph{Impact Parameter Representation}
It has to be emphasized that both of these representation for the T-matrix are not impact parameter (Fourier) representations in the strict sense, since the integrand has a dependence on both $\vecb$ and $\vecq$, the latter through 
\beq
K = k\cos\left(\frac\theta 2\right) = k\sqrt{1-\frac{\vecq^2}{4k^2}}.  
\enq
At this point, it is appropriate to mention the eikonal expansion of the T-matrix by Wallace \cite{Wallace}, which is a systematic high energy expansion of the scattering process, and possesses some properties analogous to our approach. For instance, up to second order in inverse powers of the incoming momenta $k$, it reads
\beq
T_{i\rightarrow f} = i\frac k m \int d^2b\:e^{-i\vecq\cdot\vecb}\left\{\exp \left(i\left[\chi_0(\vecb) +\tau_1(\vecb) + \tau_2(\vecb)\right] -\omega_2(\vecb)\right) -1\right\},
\enq
where the subscript accompanying each term indicates the inverse power of $k$ attached to it (while the velocity $v = k/m$ is treated as a constant). Similarly as in our representations, the eikonal approximation consists of integrating so-called scattering phases over a two-dimensional impact parameter vector $\vecb$. However, unlike ours, the eikonal representation is indeed an impact parameter representation, and the factor in front of the T-matrix is $k/m$ and not $K/m$, a difference that plays a role for large scattering angles. Since we will compare our own results to the eikonal phases later, let us mention them now, for a spherically symmetric potential, (general expressions can be found in \cite{Wallace}, or later in this work, in another form)
\beqa
\chi_0 &=& -\frac 2 v \int_0^{\infty} dZ\: V\left(\sqrt{b^2 + Z^2}\right)\label{x0w}\\
\tau_1 &=& -\frac 1 k\frac 1 {v^2}\left(1 + b\frac{d}{db}\right)\int_0^{\infty} dZ\: V^2\left(\sqrt{b^2 + Z^2}\right)\\
\tau_2 &=& -\frac 1 {k^2} \frac{1}{v^3}\left(1 + \frac53 b\frac{d}{db}+\frac13b^2\frac{d^2}{db^2}\right)\int_0^{\infty} dZ\: V^3\left(\sqrt{b^2 + Z^2}\right) -\frac{b\chi_0^{'3}}{24k^2}\\
\omega_2 &=& \frac {b}{8k^2} \chi_0'\Delta_b\chi_0=\frac{1}{16k^2}\frac1b\frac d{db}\left(b\chi_0'\right)^2 ,\label{omega2w}
\enqa
where $\Delta_b$ is the two-dimensional Laplacian acting on $\vecb$.


%% file: F-JPrinciple.tex
\newpage
\section{The Feynman-Jensen Variational Principle} \label{FJ}
The two formulae (\ref{T33}) and (\ref{T31}) contain a lot of physics. On one hand, it is possible to show (see \cite{Rosenfelder}) that, in both representations, a developpment in series of the exponential term that contains the potential leads to the Born expansion, term per term. On the other hand, in \cite{Rosenfelder} an high energy approximation of the T-matrix is also done, by expanding the phases $\chi_\vecK$, or $\chi_{ray}$, simultaneously in powers of $\vecv$ and $\vecw$. This leads to the eikonal expansion we just discussed, and eikonal-like approximations. However, and this is the topic of the thesis, these identities can also serve in a third kind as an approximation procedure, namely, one can make use of the Feynman-Jensen variational principle. This principle is based on the fact that, for a given action $S$ that depends on $x$, the quantity
\beq
\int \mathcal D x\:e^{-S} \label{paths}
\enq
can be exactly rewritten as
\beqa \nonumber
\int \mathcal D x\:e^{-S} &=& \frac{\int \mathcal D x\:e^{-(S-S_t)}e^{-S_t}}{\int \mathcal D x\: e^{-S_t}}\int \mathcal D x\:e^{-S_t}\\
&=:& \la e^{-(S-S_t)}\rat\int \mathcal Dx \:e^{-S_t}, \label{cum}
\enqa
for any other action $S_t$, which is, in practice, chosen so that one can solve the path integral on the right side of the last term. One can use then the Jensen inequality for convex functions on the term on the left side,
\beq 
\la \exp{\left[-(S-S_t)\right]}\rat \geq \exp \left[-\la S-S_t\rat\right],
\enq
so that one gains an approximation of the quantity (\ref{paths}), together with a lower bound, since the terms that enter can be readily calculated, if the action Ansatz is suitable enough. Such an approach is used, for example, in statistical mechanics, where path integrals in euclidean time show up.\newline
However, our situation does not correspond exactly to this formulation. The quantities that we have in (\ref{T33}) and (\ref{T31}) are of the form
\beq
\int \mathcal D x\:e^{iS},
\enq
with (for the eikonal representation),
\beq
\mathcal Dx = \mathcal D\vecv \mathcal D\vecw,
\enq
and
\beq
S = \frac m 2\int dt\:\left(\vecv^2(t)-\vecw^2(t)\right) +\chi_\vecK(\vecb,\vecv,\vecw).
\enq
In that case it still holds
\beqa \nonumber
\int \mathcal D x\:e^{iS} &=& \frac{\int \mathcal D x\:e^{i(S-S_t)}e^{iS_t}}{\int \mathcal D x\: e^{iS_t}}\int \mathcal D x\:e^{iS_t}\\
&:=& \la e^{i(S-S_t)}\rat\int \mathcal Dx \:e^{iS_t} \label{cum2}.
\enqa
However, the weight function over which is taken the expectation value $\la e^{i(S-S_t)}\rat$ is now a complex function , so one cannot use the Jensen inequality, and looses the lower bound information. Nevertheless, although it is not a minimum, the corresponding quantity still is stationary with respect to $S_t$: Let us define the functional
\beqa
F[S_t] &=& \nonumber\exp\left(i\la S-S_t\rat\right)\int \mathcal Dx\:e^{iS_t}\\
&=:& \exp\left(i\la S-S_t\rat\right)m_0  \label{F}
\enqa
Obviously, 
\beq
F[S] = \int \mathcal Dx\:e^{iS}
\enq 
is the quantity we are looking for. To prove stationarity, let us consider a differential change in $S_t$ given by $\delta S_t$. Then one has that
\beq \label{dF}
\left.\delta F\right|_{S_t = S} = \delta m_0 + i m_0\:\left.\delta\la S-S_t\rat\right|_{S_t = S}.
\enq
It holds further,
\beq
\delta m_0 = \delta \int \mathcal D x\:e^{iS_t} = i \int \mathcal D x\:e^{iS_t}\delta S_t,
\enq
while
\beq
\left.\delta\la S-S_t\rat\right|_{S_t = S} = \left.\delta \frac 1{m_0} \int \mathcal D x\:e^{iS_t}(S-S_t)\right|_{S=S_t} = -\frac{1}{m_0}\int \mathcal D x\:e^{iS_t}\delta S_t,
\enq
so that one effectively has in (\ref{dF})
\beq
\left.\delta F\right|_{S_t =S} = 0.
\enq
We have thus proven the stationarity of (\ref{F}) at $S$. Therefore, although one does not find a minimum or a maximum, as we said, one can hopefully still have a good approximation by a suitable choice of action $S_t$.     
\paragraph{Our Ansatz} 
The trial action $S_t$ has to be chosen so that the quantity $m_0$ and the average $\la S-S_t\rat$ can be exactly solved. This restricts the possibilities in the class of quadratic functions of their arguments. As we have mentioned at the beginning of this section, an high-energy expansion follows from expanding the scattering phases $\chi_\vecK$ and $\chi_{ray}$ in powers of $\vecv$ and $\vecw$. The effect of such an approach to first order is to render the interacting part of the action linear in $\vecv$ and $\vecw$. We will investigate in this work a simple Ansatz for the action, which is to add a linear term in $\vecv$ and $\vecw$ to the free term of $S$ that we have. Our approach is therefore similar to the high energy expansion, except for the fact that the optimization procedure will select for us not the linear term corresponding to this high energy expansion, but the best possible one. We expect thus our approach to have a greater validity than for high energies only. The trial action therefore reads
\beq
\boxed{S_t(\vecv,\vecw) = \frac{m}{2}\int dt\:\left(\vecv^2(t)-\vecw^2(t)\right) + \int dt\:\vecB(t)\cdot\vecv(t) + \int dt\:\vecC(t)\cdot\vecw(t)}
\enq
in the eikonal representation, while in the ray representation,
\beq
\boxed{S_t(\vecv,w) = \frac{m}{2}\int dt\: \left(\vecv^2(t)-w^2(t)\right) + \int dt\:\vecB(t)\cdot\vecv(t) + \int dt\:\C(t)\cdot w(t).}
\enq

\noindent The $B's$ and $\C's$ are functions that will be determined by the stationarity of $F$:
\beq \label{vareq33}
\frac{\delta F}{\delta\vecB(t)} =0 ,\quad \frac{\delta F}{\delta\vecC(t)} =0,
\enq 
respectively
\beq \label{vareq31}
\frac{\delta F}{\delta\vecB(t)} =0 ,\quad \frac{\delta F}{\delta\C(t)} =0,
\enq
and the T-matrices will then be approximated by
\beqa
T^{(3-3)}_{i\rightarrow f}&=& i\frac K m\int d^2b\:e^{-i\vecq\cdot\vecb}\left(F[S_t(\vecv,\vecw)] -1\right),\label{Tap33}\\ 
T^{(3-1)}_{i\rightarrow f}&=& i\frac K m\int d^2b\:e^{-i\vecq\cdot\vecb}\left(F[S_t(\vecv,w)]-1\right).\label{Tap31}
\enqa
The corresponding expectation values that enter $F$ and need to be computed are
\beqa
\la(S-S_t)\rat &=&\la\chi_\vecK\rat  -\la\Bv\rat -\la\Cw\rat \label{exp33}\\
\la(S-S_t)\rat_{ray} &=& \la\chi_{ray}\rat - \la\Bv\rat - \la(\C, w)\rat,\label{exp31}
\enqa
where we introduced the short-hand notation
\beq
(\mathbf f,\mathbf g) := \int dt\:\mathbf f(t)\cdot\mathbf g(t),
\enq
that we will use frequently from now on.
\paragraph{Corrections}
Corrections to this approximation can be gained by recognizing that the quantity
\beq
\exp\left[i\la S-S_t \rat\right]\nonumber
\enq
is the first term in an expansion in cumulants of 
\beq
\la e^{i(S-S_t)}\rat,\nonumber
\enq
entering equation (\ref{cum2}). The cumulants $\lambda_k$ of the distribution are defined \cite{Abramowitz} through
\beq
\la e^{it(S-S_t)} \rat = \exp\left[\sumi k1\infty \frac{(it)^k} {k!}\lambda_k\right].
\enq
On the other hand, the expansion in moments
\beq
\la e^{it(S-S_t)}\rat = \sumi k0\infty \frac{(it)^k}{k!}\la(S-S_t)^k\rat,
\enq
permits, through comparison of powers of $t$, to express the cumulants in terms of the moments. One has for the first two,
\beqa
\lambda_1 &=& \la (S-S_t)\rat \\
\lambda_2 &=& \la (S-S_t)^2\rat - \la (S-S_t)\rat^2.  \label{correction}
\enqa
In this work we will in due time consider the first correction $\lambda_2$, with the corresponding functional
\beq
F[S_t] = m_0 \exp\left(i\la S-S_t\rat \right)\exp\left(-\frac 12 \left[\la (S-S_t)^2\rat - \la (S-S_t)\rat^2\right]\right).
\enq

%% file: EikonalRepresentation.tex
\newpage
\section{Variational Approximation in the Eikonal Representation}\label{EikonalRepresentation}
In this section, we apply the Feynman-Jensen variational principle to the eikonal representation (\ref{T33}) of the T-matrix, and investigate its implications. The procedure is the following:
\begin{itemize}
	\item First, we perform the computation of all needed quantities, in terms of the trial functions $\vecB$ and $\vecC$. These are the expectation values (\ref{exp33}) that build up the first cumulant $\lambda_1$, and the normalisation factor $m_0$.
	\item Second, we impose the variational equations (\ref{vareq33}) to the trial functions.
	\item Third, we read out from those the scattering phases and the T-matrix.  
	\item Fourth, we obtain the first correction, which is the second cumulant $\lambda_2$,
	\item And fifth, we have a look at the high-energy behavior of our results. 
\end{itemize}
\paragraph{First Cumulant.}
The first cumulant $\lambda_1$ is the first central moment $\la S-S_t\rat$. It leads to the following approximation
\beq
T^{(3-3)}_{i\rightarrow f} = i\frac K m \int d^2b\:e^{-i\vecb\cdot\vecq}\:T_1^{(3-3)}(\vecb,\vecK),
\enq
with
\beqa \nonumber 
T_1^{(3-3)}(\vecb,\vecK)  = F[S_t] -1  &=& m_0\exp\:i\la S-S_t\rat -1 \\
&=&m_0\exp\:i\la\chi_\vecK\rat\exp\left(-i\la\Bv+\Cw\rat\right) - 1. \label{firstcumulant}
\enqa
The quantity $m_0$ is the normalisation factor
\beqa
\int \mathcal D\vecv \mathcal D\vecw \:e^{iS_t} &=& \int \mathcal D\vecv \mathcal D\vecw \exp\lp i\frac m2 \left[(\vecv,\vecv)-(\vecw,\vecw)\right] + i\Bv + i\Cw \rp \nonumber\\
&=&\exp\left(-\frac{i}{2m}\left[\left(\vecB,\vecB\right)-\left(\vecC,\vecC\right)\right]\right)
\enqa
Let us consider first the term in (\ref{firstcumulant}) that involves the functions $\vecB$ and $\vecC$. This expectation value can be written as
\beqa
&&\la\Bv+\Cw\rat = \frac{1}{m_0}\intvw\left[\Bv+\Cw\right]\exp\left[iS_t(\vecv,\vecw)\right]\nonumber\\ 
&=&\nonumber \frac{-i}{m_0}\left.\frac{d}{da}\right|_{a =1}\intvw\exp\:\left(i\int \frac {m}{2} \left(\vecv^2-\vecw^2\right) +ia\left[\Bv+\Cw\right]\right).
\enqa
This last expression is simply the normalisation constant $m_0$, with the trial functions multiplied by the additional factor $a$. One obtains
\beqa
\la\Bv+\Cw\rat &=& -\frac{i}{m_0}\left.\frac{d}{da}\right|_{a =1}\exp\left(-\frac{i}{2m}a^2\left[\left(\vecB,\vecB\right)-\left(\vecC,\vecC\right)\right]\right)\\
&=&- \frac{1}{m}\left[\left(\vecB,\vecB\right)-\left(\vecC,\vecC\right)\right].
\enqa
Therefore, formula (\ref{firstcumulant}) for the first term in the cumulant expansion becomes
\beq \label{T1}T_1^{(3-3)}(\vecb,\vecK) = \exp\left[i\la\chi_\vecK\rat\right]\exp\left\{\frac i {2m}\left[\left(\vecB,\vecB\right)-\left(\vecC,\vecC\right)\right]\:\right\}-1.
\enq
The first cumulant is thus given by
\beq\label{l133}
\cumul1 = \la\chi_\vecK\rat+\frac 1 {2m}\left[\left(\vecB,\vecB\right)-\left(\vecC,\vecC\right)\right]
\enq
The variational equations are now easily read out. Since we want our expression (\ref{T1}) to be stationary with respect to the trial functions, it must hold
\beq \label{varB}\vecB(t) = -m\:\frac{\delta\la\chi_\vecK\rat(\vecB,\vecC)}{\delta\vecB(t)},\enq
respectively
\beq \label{varC}\vecC(t) = m\:\frac{\delta\la\chi_\vecK\rat(\vecB,\vecC)}{\delta\vecC(t)}.\enq
\paragraph{Computation of $\la\chi_\vecK\rat$.}
This is done most easily by Fourier expanding the potential function,
\beq \la\chi_\vecK\rat = -\frac 1 {m_0}\intvw\int \frac{d\vecp}{(2\pi)^3}\widetilde V(\vecp)\int dt\:e^{iS_t}\:e^{\:i\vecp
\left[\vecb +\frac \vecK m t + \vecx_v(t)-\vecx_w(0)\right]}.
\enq
Since $\vecx_v$ and $\vecx_w$ are linear functions of $\vecv$, $\vecw$ respectively, the functional integrations can indeed be exactly done. Let us first have a look at the $\vecv$ integration :
\beqa \nonumber
&&\int \mathcal D \vecv \exp\left(-i\frac{m}{2}\left(\vecv,\vecv\right) + i\Bv +i\frac{1}{2}\vecp\int ds\:\vecv(s)\sgn(t-s)\right)\\
&=& \lim_{T\rightarrow \infty}\exp\left(-\frac{i}{2m}\int_{-T}^{+T} ds\:\left[\vecB(s)+\frac 1 2 \vecp\:\sgn(t-s)\right]^2\right).
\enqa
The $\vecw$ integration is similar and results in
\beq
\lim_{T\rightarrow \infty}\exp\left(\frac{i}{2m}\int_{-T}^{+T} ds\:\left[\vecC(s)-\frac 1 2 \vecp\:\sgn(-s)\right]^2\right).
\enq
From the terms in the squared brackets, the $\vecB^2$ and $\vecC^2$ terms cancel against the normalisation constant. Both terms that are seemingly infinite cancel each other, and we are left with 
\beq \nonumber \la\chi_\vecK\rat = -\int dt\:V\left(\vecb + \Kom t -\frac{1}{2m}\left[\int ds\: \vecB(s)\sgn(t-s) + \vecC(s)\sgn(-s) \right]\right).
\enq 
From now on we will write $\vecx(t)$ for the trajectory
\beq \nonumber \vecb + \Kom t -\frac{1}{2m}\left[\int ds\: \vecB(s)\sgn(t-s) + \vecC(s)\sgn(-s) \right] =: \vecx(t).\enq
Our result is thus
\beq\la\chi_\vecK\rat = -\int dt\:V(\vecx(t)).\enq
\subsection{Variational Equations and the Variational Trajectory}
We see now clearly that the role of the trial functions $\vecB$ and $\vecC$ is to determine a better reference path that the particle travels along.
As expected, setting $\vecB$ and $\vecC$ to zero reduces to the straight line trajectory, and subsequently to the eikonal approximation, here in the variant of Abarbanel and Itzykson (AI) \cite{Abarbanel}, where $K$ appears everywhere in place of $k$. We will see that in our case too, where the choice of trajectory follows from the variational equations (\ref{varB}) and (\ref{varC}), the trajectory can be readily interpreted. The variational equations translate to
\beqa \label{VarB2}
\vecB(t) &=& -\frac 1 2 \int ds\:\grads\sgn(s-t),\\
\vecC(t)&=& \:\:\:\frac 1 2 \int ds\:\grads \mathrm \sgn(-t), \label{VarC2}
\enqa
which, with the help of the identity
\beq
-\int ds\:\left[\mathrm{sgn}(s-t)\mathrm{sgn}(s-t') - 1 \right] = 2\left|t-t'\right|,
\enq
implies in turn
\beq\boxed{\label {eiktraj}\vecxt = \vecb + \Kom t -\frac{1}{2m}\int ds\:\nabla V(\vecx(s))|t-s|.}\enq
This identity is none other than a integral representation of a solution $\vecxt$ to the Newtonian equation of motion for a point particle. Indeed, the first two terms on the right hand side are the most general free solution, and the third term is the convolution of the gradient of the potential function with the Green function of that equation of motion. This is formally seen by applying twice a time derivative on both sides of the equation. Since 
\beq
\frac{d}{dt}|t-s| = \mathrm {sgn}(t-s)\quad\mathrm{and}\quad \frac{d}{dt}\mathrm {sgn(t-s)} = 2\delta(t-s), 
\enq
the computation is straightforward :

\beq \label{classic}
m\ddot \vecx (t) = -\gradt.
\enq
Interestingly enough, in this approximation of quantum mechanical scattering theory, an important feature of which is the scattering integral equation, its classical counterpart appears. \newline
\paragraph{Asymptotic Behavior.}
In order to get a feeling for the role played by the two fixed parameters $\vecb$ and $\vecK$ in this trajectory, let us consider more closely its asymptotic behavior. We will always assume in the subsequent considerations that the potential falls off rapidly enough so that the integrals involved converge. For very large positive times $t$, the term $t-s$ in the integral can be considered always positive. One can then rewrite the trajectory as
\beq \label{L_f}
\vecxt \stackrel{t \rightarrow \infty}{\rightarrow} \vecb_f + t\vecv_f,
\enq
where the starting point $\vecb_f$ and asymptotic final velocity $\vecv_f$ are related to $\vecb$ and $\vecK$ by
\begin{eqnarray}
\vecb_f &:=& \vecb + \frac{1}{2m}\int ds\:\nabla V(\vecx(s))s\quad := \vecb + \frac12\bar\vecb\\
\vecv_f &:=& \Kom -\frac{1}{2m}\int ds\:\nabla V(\vecx(s))\quad:= \Kom + \frac\vecQ{2m}.
\end{eqnarray}
In the very same vein, one obtains for very large negative times
\beq\label{L_i}
\vecxt \stackrel{t \rightarrow -\infty}{\rightarrow} \vecb_i + t\vecv_i,
\enq
with
\begin{eqnarray}
\vecb_i &=& \vecb - \frac12\bar\vecb\\
\vecv_i &=& \Kom - \frac\vecQ{2m}.
\end{eqnarray}
These relations implies immediately
\begin{eqnarray}
\vecb &=& \frac{1}{2}(\vecb_i +\vecb_f)\\
\Kom &=&\frac{1}{2}(\vecv_i +\vecv_f). \label{K}
\end{eqnarray}
One deduces from these relations that the integral equation for $\vecx(t)$ is (at least formally) equivalent to the boundary value problem given by the Newtonian differential equation of motion, with the boundary conditions determined through
\beqa
\vecb &=& \frac 12 \:\lim_{t\rightarrow\infty}\left\{ \vecx(t)+\vecx(-t) - t\left[\dot\vecx(t) -\dot\vecx(-t)\right]\:\right\}\\
\frac \vecK m &=& \frac 12 \:\lim_{t\rightarrow\infty}\left\{\dot\vecx(t) +\dot\vecx(-t)\right\}.
\enqa
Equation (\ref{K}) says that $\vecK$ keeps very nicely its meaning of mean momentum, something that had \textsl{a priori} no reason to be expected from the variational principle. The quantity $\vecQ$, on the other hand, is then the momentum transfer due to the deflection in this trajectory only, but cannot be compared with the momentum transfer $\vecq$ of the scattering process we are investigating. Since the energy is conserved along the trajectory, $\vecQ$ is a purely transversal quantity. The interpretation of $\vecb$ for a general potential is on the contrary more difficult. Sure is that if $\vecb$ is large enough, the trajectory is nothing more than the straight eikonal line. This follows from the fact that if the potential at $\vecb$ and beyond is negligible, then this straight trajectory is a solution to the integral equation. Nevertheless, since one integrates over $\vecb$ to obtain the scattering amplitude, we see that every classical trajectory with mean momentum $\vecK$ contributes in our approximation to the scattering process. It is very similar in spirit with the eikonal expansion, where all straight line trajectories with velocity $\veck/m$ are taken into account.
\paragraph{Constants of Motion.}
The asymptotic conditions we derived above permit us to derive simple expressions for the energy and, in the case of a potential with rotational symmetry, for the angular momentum of this classical curve. According to (\ref{L_i}) and (\ref{L_f}), the angular momentum $m\vecx \wedge \dot \vecx$ at infinity is given by
\beq 
\vecL = m\vecb_f\wedge\vecv_f = m\vecb_i\wedge\vecv_i.
\enq
It is possible to combine these two to obtain
\begin{eqnarray}
\vecL &=& \frac{m}{2}(\vecb_i\wedge\vecv_i + \vecb_f\wedge\vecv_f) \\
  &=&\frac{m}{2}\left(\left(\vecb-\frac12\bar\vecb\right)\wedge\left(\Kom-\frac\vecQ{2m}\right) +\left(\vecb+\frac12\bar\vecb\right)\wedge\left(\Kom+\frac\vecQ{2m}\right)\right)\\
  &=&\vecb\wedge\vecK + \frac14\bar\vecb\wedge\vecQ
\end{eqnarray}
The same procedure applied to the energy gives as a result
\beq \label{energy}
E = \frac{\vecK^2}{2m} + \frac{\vecQ^2}{8m},
\enq
These two relations prove to be useful for instance if one is interested in a high $K$ expansion.
\paragraph{Symmetries.}
Under some assumptions on the potential, the trajectory exhibit some further symmetries.
Let us suppose that the potential function is such that it depends only on the modulus of the components that are perpendicular and parallel to the mean momentum,
\beq
V(\vecx) = V(\vecx_\parallel^2, \vecx_\perp^2).
\enq
To this class of functions belong of course the potentials that possess rotational symmetry, but not only those, since each component could be treated differently, like in a "cylindrically" symmetric potential. Although this last category may look a bit artificial, it has some importance, since we will encounter such a function in the ray representation of the T-matrix, where the perpendicular and longitudinal parts are handled in different ways. Anyway, for such potentials it clearly holds that
\beq \label{cond}
\nabla_\perp V(\vecx) = f(\vecx_\parallel^2, \vecx_\perp^2) \vecx_\perp\quad \mathrm{and}\quad \nabla_\parallel V (\vecx)= g(\vecx_\parallel^2, \vecx_\perp^2) \vecx_\parallel
\enq
for some scalar valued functions $f$ and $g$ that may or may not differ. We claim that this implies in turn the symmetry properties
\beq
\vecx_\perp (-t) = \vecx_\perp(t) \quad\mathrm{and}\quad \vecx_\parallel(-t) = -\vecx_\parallel(t).
\enq
These properties, together with the conditions (\ref{cond}) we stated above, are clearly seen to be compatible with the defining equation (\ref{eiktraj}). Although physically intuitive, to affirm that $\vecx(t)$ must possess these, one can argue as follows: the standard way to approach the solution to such an integral equation is to proceed  through iteration. Starting with
\beq
\vecx_0(t) := \vecb + \frac \vecK m t,
\enq
one defines recursively
\beq
\vecx_n(t) := \vecb + \frac{\vecK}{m}t -\frac{v}{2K}\int ds\:\nabla V(\vecx_{n-1}(s))|t-s|,
\enq
with the velocity $v = K/m.$ 
In a high energy limit, i.e. fixed velocity but high $K$, the so-defined $\vecx_n(t)$ are the successive approximations to $\vecx(t)$ up to order $K^{-n}$. Since $\vecx_0(t)$ possesses the symmetry properties stated above, by induction over $n$ all terms satisfy them. Since such symmetry properties cannot depend on the very magnitude of $\vecK$, $\vecx(t)$ must in general obey these, too. \newline This has as a consequence, for example, as can be easily checked, that $\vecx(0)$ is purely perpendicular, and $\dot\vecx(0)$ purely longitudinal, implying, if we denote $|\vecx(t)|$ with $r(t)$,
\beq 
\left.\frac{d}{dt}r(t)\right|_{t=0} = \frac{1}{r(0)}\vecx(0)\cdot\dot\vecx(0) = 0.
\enq
In other words, $\vecx(0)$ is the closest approach to the scattering center.

\subsection{The Scattering Phases}
We have computed in the last section the expression for the T-matrix in first order in the cumulant expansion, in terms of $\la\chi_\vecK\rat$, in equation (\ref{T1}). When $\vecB$ and $\vecC$ do satisfy the variational equations (\ref{VarB2}) and (\ref{VarC2}), it becomes

\beq T_1^{(3-3)}(\vecb,\vecK) = \exp  \left(iX_0 + iX_1\right) -1\enq
where we introduced a bit of notation :

\beq \label {X033} \boxed{X_0(\vecb,\vecK) =  -\int dt\: V(\vecxt) = \la\chi_\vecK\rat}\enq

\noindent and \beq\boxed{ X_1(\vecb,\vecK) = -\frac 1 {4m}\int dt\int ds\:\gradt\cdot\grads |t-s|.}\enq
For potentials with rotational symmetry, since $\vecb$ is perpendicular to $\vecK$ one has
\beq
T_1^{(3-3)}(\vecb,\vecK) = T_1^{(3-3)}(b,K).
\enq
A first, very rough estimate of $X_0$ for small values of b can be done as follows: we denote the scale over which the potential takes on sensible values, with average value $V_0$, by $R$; since the particle following this trajectory would cross this region in a time $\approx R/v$, where $v$ is the mean velocity $K/m$, one has roughly
\beq 
X_0 \approx -V_0\frac Rv = -\epsilon\: KR, \quad \epsilon \equiv V_0/Kv.
\enq
In the last equality we have introduced the dimensionless quantity $\epsilon$, which describes the relative importance of the potential energy with respect to the kinetic energy. A similar argument in the case of $X_1$ shows
\beq
X_1 \approx \frac aK\left(\frac{\Delta V}{v}\right)^2,
\enq
where $a$ stands for the scale over which the potential varies appreciably, by an amount $\Delta V$. In the case where the quantities $a$ and $R$ are essentially the same, one has
\beq
X_1 \approx -\epsilon^2\:KR.
\enq
Of course, $a$ and $R$ can sometimes differ considerably, as in the case of a Woods-Saxon potential, that looks like a well with rounded edges. In that particular case,
\beq
X_1 \approx -\epsilon^2\:Ka.
\enq
The larger the energy of the incoming particle, the smaller epsilon, the more dominant is $X_0$ with respect to $X_1$.\newline
These two phases are very similar in shape to those of the eikonal expansion, up to first order in inverse momentum, except apparently for the minus sign in front of $X_1$. The first two phases of the eikonal expansion of Wallace read for a general potential
\beq
\int dt\:V\left(\vecb + \hat\vecK\frac k m t \right),\quad +\frac 1{4m}\int dt\:\int ds\:\nabla V\left(\vecb + \hat\vecK\frac k m t \right)\cdot\nabla V\left(\vecb + \hat\vecK\frac k m s \right)|t-s|.\enq
We will nevertheless show later that they coincide in the high energy limit.\newline
However, the two scattering phases $X_0$ and $X_1$ satisfy an amusing relation that has no eikonal counterpart. It is indeed possible to show that
\beq \label{rel}
X_0 - X_1 = \vecK\cdot(\vecb_f-\vecb_i) = \vecK\cdot\bar\vecb,  
\enq
where we have used the notation introduced for the asymptotic properties of the trajectory in (\ref{L_f}) and (\ref{L_i}). The proof reads as follows:
Considering the defining equation (\ref{eiktraj}) and its classical property (\ref{classic}), one can first rewrite $X_1$ as
\beqa
X_1 &=& \label {X133}\frac 12 \int dt\:\nabla V(\vecx(t))\cdot\left[\vecx(t) -\vecb -\frac \vecK m t\right] \\
&=&-\frac m2 \int dt\:\:\ddot\vecx(t)\cdot\left[\vecx(t) -\vecb -\frac \vecK m t\right].
\enqa
Using finite bounds for the integration interval, one can integrate by parts and consider each term on its own.
\beq \left.X_1\right|_{T} = -\frac m2 \left.\:\dot\vecx(t)\cdot\left[\vecx(t) -\vecb -\frac \vecK m t\right]\right|_{t=-T}^{t=+T} + \frac m 2 \int_{-T}^{+T} dt\:\dot\vecx^2(t) -\left.\frac \vecK m \vecx(t)\right|_{t = -T}^{t = + T} .
\enq
The second term on the right is simply
\beq
 \frac m 2 \int_{-T}^{+T} dt\:\dot\vecx^2(t) = \int_{-T}^{+T} dt\:\left[E-V(\vecx(t))\right] = 2TE + \left.X_0\right|_{T}.
\enq
By inserting the asymptotic trajectory into the two other terms, one sees after a small calculation using (\ref{energy}), that all dependence on $T$ disappears, and what is left is the formula (\ref{rel}) claimed above.\newline

\subsection{First Correction to the Variational Approximation}
In this section, we want to go a step further and compute the second order term $\lambda_2$ in the cumulant expansion of 
$\la e^{i\Delta S}\rat$. However, to keep things simple we will stick to the solutions of the variational principle we derived earlier. Especially, we expect the appearance of a real term. The main results of this rather long and technical section are equation (\ref{lambda2_}) combined with (\ref{l1var}), and equation (\ref{lambda2}).
\newline\newline
This second order term is the second central moment, or variance (see (\ref{correction})). It is related to the first two moments by
\beq
\lambda_2 = \la\Delta S^2\rat -\expd^2.
\enq
This means that our approximation of the integrand in the T-matrix in is going to be
\beq
T^{(3-3)}_{i\rightarrow f} = i\frac K m \int d^2b\:e^{-i\vecb\cdot\vecq}\:T_2^{(3-3)}(\vecb,\vecK),
\enq
with
\beqa
T_2^{(3-3)}(\vecb,\vecK) &=& m_0\exp\left(i\lambda_1-\frac 12 \lambda_2\right)-1.
\enqa
As usual, $m_0$ stands for the normalization factor
\begin{eqnarray} m_0 &=& \int\mathcal D\vecv\mathcal D\vecw\:e^{iS_t}\\
&=& \exp\left\{-\frac{i}{2m}\left[(\vecB,\vecB)-(\vecC,\vecC)\right]\right\}. 
\end{eqnarray}
To this aim, the six terms entering
\begin{eqnarray*}
\la\Delta S^2\rat  &=& \la\left(\chi_\vecK -\left(\vecB,\vecv\right)-\left(\vecC,\vecw\right)\right)^2\rat \\
&=& \la\chi_\vecK^2\rat + \la\left(\vecB,\vecv\right)^2\rat +\la\left(\vecC,\vecw\right)^2\rat\\
&-&2\la\chi_\vecK\left(\vecB,\vecv\right)\rat - 2\la\chi_\vecK\left(\vecC,\vecw\right)\rat + 2\la\left(\vecB,\vecv\right)\left(\vecC,\vecw\right)\rat
\end{eqnarray*}
need to be calculated. The last term causes no worries : since $\vecv$ and $\vecw$ are uncorrelated, the expectation value factorize out, and the term will cancel its corresponding part in $\la\Delta S\rat^2$ in $\lambda_2$.\newline
The four terms involving the trial functions $\vecB$ and $\vecC$ can be computed using a similar procedure we did when we calculated the first cumulant: we replace $\vecB$, or $\vecC$ respectively, with the help of a fictitious variable $a$ in the action $S_t$ by $a\vecB$, or $a\vecC$, and derivation of $m_0$ with respect to $a$ brings down each time a factor $i(\vecB,\vecv)$. More concretely, we have for instance, for each positive integer $n$,

\begin{eqnarray}
&&\la\Bv^n\rat = \frac{1}{m_0}\int\mathcal D\vecv\mathcal D\vecw\:\Bv^n\:e^{i\frac{m}{2}\int dt\left(\vecv^2-\vecw^2\right) +i\Bv+i\Cw}\\ 
&=&\frac{1}{m_0}\left(\frac{1}{i}\right)^n\left.\frac{d^n}{da^n} \right |_{a = 1}\int\mathcal D\vecv\mathcal D\vecw\:
e^{i\frac{m}{2}\int dt\left(\vecv^2-\vecw^2\right) +ia\Bv+i\Cw}\\
&=&\frac{1}{m_0}\left(\frac{1}{i}\right)^n\left.\frac{d^n}{da^n} \right |_{a = 1}
\exp\left\{-\frac{i}{2m}\left[a^2(\vecB,\vecB)-(\vecC,\vecC)\right]\right\}. 
\end{eqnarray}
An even more concise form can be written after recognizing that this very last expression represents a rescaled Hermite polynomial. Indeed, we obtain 
\beq
\la\Bv^n\rat = i^n\left(\frac{i}{m}(\vecB,\vecB)\right)^{n/2}He_n\left(\sqrt{\frac{i}{m}(\vecB,\vecB)}\right),
\enq
where $He_n(x)$ is the Hermite polynomial of degree $n$, defined \cite{Abramowitz} by
\beq He_n(x) = (-1)^n\:e^{x^2/2}\frac{d^n}{dx^n}\:e^{-x^2/2}.\enq
In a completely analogous way, one gets
\beq
\la\Cw^n\rat = i^n\left(-\frac{i}{m}(\vecC,\vecC)\right)^{n/2}He_n\left(\sqrt{-\frac{i}{m}(\vecC,\vecC)}\right).
\enq
Since $He_2(x) = x^2 -1$, it follows immediately that
\begin {eqnarray}
\la\Bv^2\rat &=& \frac{i}{m}(\vecB,\vecB)+\frac{1}{m^2}(\vecB,\vecB)^2\\ \label{BV^2}
&=&\frac{i}{m}(\vecB,\vecB)+\la\Bv\rat^2,
\end {eqnarray}
respectively
\begin {eqnarray}
 \la\Cw^2\rat &=&- \frac{i}{m}(\vecC,\vecC)+\frac{1}{m^2}(\vecC,\vecC)^2\\ \label{CW^2}
&=&-\frac{i}{m}(\vecC,\vecC)+\la\Cw\rat^2.
\end {eqnarray}
In a similar vein,
\begin{eqnarray}
\la\chi_\vecK\Bv\rat &=& \frac{1}{m_0}\int\mathcal D\vecv\mathcal D\vecw\: \chi_\vecK \Bv e^{iS_t}\\
&=&-i\frac{1}{m_0}\left.\frac{d}{da}\right|_{a=1}\int\mathcal D\vecv\mathcal D\vecw \:\chi_\vecK\:e^{iS_t^a}\\
&=&-i\frac{1}{m_0}\left.\frac{d}{da}\right|_{a=1} m_0^a\la\chi_\vecK\rat^a.\\
&=& \la\chi_\vecK\rat\la\Bv\rat -i\left.\frac{d}{da}\right|_{a = 1}\la\chi_\vecK\rat^a
\end{eqnarray}
In the last three lines, the upper script $^a$ denotes the usual quantity with $\vecB$ replaced by $a\vecB$. We know already what $\la\chi_\vecK\rat$ is from our study of the first cumulant:
\begin{equation*}
\la\chi_\vecK\rat^a = -\int dt\: V\left(\vecb + \frac{\vecK}{m}t -\frac{1}{2m}\int ds\left[a\vecB(s)\:\mathrm{sgn}(t-s)+\vecC(s)\:\mathrm{sgn}(-s)\right]\right).
\end{equation*}
It follows then
\beq \label{chiBV}
\la\chi_\vecK\Bv\rat = \la\chi_\vecK\rat\la\Bv\rat -\frac{i}{2m}\int dt\int ds\:\nabla V(\vecx(t))\cdot\vecB(s)\mathrm {sgn}(t-s),
\enq
and in an analogous way,
\beq \label{chiCW}
\la\chi_\vecK\Cw\rat = \la\chi_\vecK\rat\la\Cw\rat -\frac{i}{2m}\int dt\int ds\:\nabla V(\vecx(t))\cdot\vecC(s)\mathrm {sgn}(-s).
\enq
\noindent We now gather equations (\ref{BV^2}), (\ref{CW^2}), (\ref{chiBV}), (\ref{chiCW}) to obtain
\begin{eqnarray}
\lambda_2 &=& \la\Delta S^2\rat -\expd^2\\
&=& \nonumber \la\chi_\vecK^2\rat -\la\chi_\vecK\rat^2 +\frac{i}{m}(\vecB,\vecB) -\frac{i}{m}(\vecC,\vecC)\\
&&+\:\frac{i}{m}\int dt\int ds\:\nabla V(\vecx(t))\cdot\left[\vecC(s)\sgn(-s)+\vecB(s)\sgn(t-s)\right]
\end{eqnarray}
If $\vecB$ and $\vecC$ are taken to satisfy the variational equations we derived earlier, we obtain the representation
\beqa \nonumber\lambda_2 &=& \la\chi_\vecK^2\rat -\la\chi_\vecK\rat^2 +\frac{i}{2m}\int dt\int ds\:\nabla V(\vecx(t))\nabla V(\vecx(s))\left|t-s\right|.\\ \label{l1var}
&=& \la\chi_\vecK^2\rat -\la\chi_\vecK\rat^2  -2iX_1\label{l233}\enqa 
As expected, divergent phases again cancel with the help of the antivelocity. Therefore, it holds
\begin{eqnarray}
T_2^{(3-3)}(\vecb,\vecK) &=&  m_0\:\exp\left(i\lambda_1-\frac{1}{2}\lambda_2\right)-1\\
&=& \nonumber\exp \left(iX_0 +2iX_1-\frac{1}{2}\left[\la\chi_\vecK^2\rat - X_0^2\right]\right)-1
\end{eqnarray}
We are thus left with the task to compute $\la\chi_\vecK^2\rat$. 
\paragraph{Computation of $\la\chi_\vecK^2\rat$.}

Again, this can be can be done by Fourier expanding the potential, and solving the $\vecv$,$\vecw$ functional integrals :
\begin{eqnarray}
\la\chi_\vecK^2\rat &=& \frac{1}{m_0}\int\mathcal D\vecv\mathcal D\vecw\:e^{iS_t}\left[-\int dt\:V\left(\vecb +\frac{\vecK}{m}t + \vecx_v(t)-\vecx_w(0)\right)\right]^2\\
 \nonumber &=&\frac{1}{m_0}\int dt_1\int dt_2\int\dpun\int\dpdeux\int\mathcal D\vecv\mathcal D\vecw\:e^{iS_t}\:\widetilde V(\vecp_1)\widetilde V(\vecp_2)\\
 &&\cdot\exp\left\{i\sum_{i=1}^2\vecp_i\left[\vecb+\Kom t_i+\vecx_v(t_i)-\vecx_w(t_i)\right]\right\}.
 \end{eqnarray} 
Let us consider first the $\vecv$ integration.
\beqa
\nonumber && \int\mathcal D\vecv \exp \left[i\frac{m}{2}(\vecv\cdot\vecv)+i\Bv+\frac i2 \sum^{2}_{i=1}\vecp_i\int ds\: \vecv(s)\sgn(t_i-s)\right] \\
&=&\exp\: -\frac i {2m}\int ds\left(\vecB(s)+\frac 1 2 \sum_{i = 1}^2\vecp_i\:\sgn(t_i-s)\right)^2.
\enqa
Similarly, the $\vecw$ integration results in
\beq
\exp\left\{ \frac i {2m}\int ds\left(\vecC(s)+\frac 1 2 \sum_{i = 1}^2\vecp_i\:\sgn(-s)\right)^2\right\}.
\enq
We see once again that the antivelocity cancels the divergent phases. From the squared terms, $\vecB^2$ and $\vecC^2$ are killed by the normalisation constant, and the others combine to give the result
\beqa \nonumber
\la\chi_\vecK^2\rat &=& \int dt_1\int dt_2\int\dpun\int\dpdeux\:\widetilde V(\vecp_1)\widetilde V(\vecp_2)\\ \label{chi2}
&&\cdot\exp\left\{i\left[\vecp_1\cdot\vecx(t_1)+\vecp_2\cdot\vecx(t_2)\right]\exp\left(\frac i{2m}\vecp_1\cdot\vecp_2\left|t_1-t_2\right|\right)\right\}.
\label{lambda2_}
\enqa
It is also possible to integrate over the momenta to obtain a real space representation. To this aim, we first rewrite it as
\beq
\la\chi_\vecK^2\rat=\int dt_1\int dt_2\int\frac{d\vecp}{(2\pi)^3}\:\widetilde V(\vecp)V\left(\vecx(t_1)+\frac{1}{2m}\vecp|t_1-t_2|\right)e^{i\vecp\cdot\vecx(t_2)},
\enq
and expand the potential function in a power series,
\beq 
\la\chi_\vecK^2\rat = 
\sum_{|\alpha|\geq 0}\frac 1 {\alpha!}\int dt_1\int dt_2\int\frac{d\vecp}{(2\pi)^3}\: \vecp^\alpha \widetilde V(\vecp)\left(\frac{|t_1-t_2|}{2m}\right)^{|\alpha|}D^{\alpha}V\left(\vecx(t_1)\right)e^{i\vecp\cdot\vecx(t_2)}.
\enq
It suffices then to note that $\vecp^\alpha\widetilde V(\vecp) = (-i)^{|\alpha|} \widetilde{D^\alpha V}(\vecp)$ to conclude
\beq\label{chi2_}
\la\chi_\vecK^2\rat = \sum_{|\alpha| \geq 0}\frac{1}{\alpha !}\left(\frac {-i} {2m}\right)^{|\alpha|}\int dt_1\int dt_2\: D^\alpha V(\vecx(t_1)) D^\alpha V(\vecx (t_2))\:|t_1-t_2|^{|\alpha|}.\enq

Obviously, the first term in this infinite sum is $\la\chi_\vecK\rat^2$, and those with $|\alpha| = 1$ build twice the phase $X_1$. Therefore, the second cumulant can also be written as
\beq \label{lambda2}\lambda_2 = \sum_{|\alpha| \geq 2}\frac{1}{\alpha !}\left(\frac {-i} {2m}\right)^{|\alpha|}\int dt_1\int dt_2\: D^\alpha V(\vecx(t_1)) D^\alpha V(\vecx (t_2))\:|t_1-t_2|^{|\alpha|},\enq
where the sum starts at $|\alpha| = 2$ only.\newline
While this form is useful for a high energy expansion, it is, on the other hand, rather unpractical for numerical computations.

\subsection{High Energy Expansion}
It is interesting to investigate the large energy behavior of $T_{i\rightarrow f}^{(3-3)}$, so that it can be then immediately compared to the systematic expansion done by Wallace in \cite{Wallace}. In such an approach, all quantities are expanded in inverse powers of $k$, while the velocity term $k/m$ is treated as fixed, as appropriate for a high energy situation. Starting from the phases $X_0$ and $X_1$ and the second cumulant we just derived, we will investigate the first and second order of the high energy expansion, and obtain the corresponding T-Matrices, defined through
 \beq \label{T_I}
 T_{I}^{(3-3)}(\vecb) = \exp\lp i \left[\chi_0(\vecb) + \chi_1(\vecb)\right]\rp -1,
 \enq
and
\beq \label{T_II}
T_{II}^{(3-3)}(\vecb) = \exp\lp i \left[\chi_0(\vecb) + \chi_1(\vecb) + \chi_2(\vecb)\right]-\omega_2(\vecb)\rp -1.
 \enq
The subscripts under the different terms indicates the power of $1/k$ attached to it in the $1/k$ expansion. 
In doing so, one has to be careful with the corrections arising from the fact that it is not the initial momentum $k$, but the mean momentum $K$, that appears in our path-integral formulation of the T-matrix. However, since
\beqa \nonumber
K &=& k\cos\lp \frac \theta2\rp = k\:\sqrt{1-\frac{q^2}{4k^2}}\\
&=&  k\lp 1 - \frac{q^2}{8k^2} - \frac{q^4}{128k^4} + \cdots\rp ,
\enqa
these corrections will not appear at first order. This is why notably no imaginary phase $\omega_1(\vecb)$ can appear at all at first order . We will see later how to handle these corrections.
\paragraph{First Order.} 
The leading term of the trajectory $\vecx(t)$ in a high $K$ expansion is given by
\beq
\vecx_0(t) := \vecb + \frac\vecK m t.
\enq
The first order contribution is obtained in the following way:

\beqa
\vecx(t)\nonumber &=& \vecb + \frac\vecK m t -\frac{1}{2m}\int ds\:\grads|t-s|\\
&=&\vecx_0(t) - \frac{v}{2K}\int ds \: \nabla V\left(\vecx_0(s)\right)|t-s| + \cdots,
\enqa
where we wrote $v$ for $K/m$.
This implies in turn that the leading phase $\xzero$ can be written as
\beqa
\xzero \label{chi_0} &=& -\int dt\: V(\vecx(t))= -\int dt\:V\lp\vecx_0(t)\rp +\nonumber\\
&+& \frac{v}{2K}\int ds\:\int dt\: \nabla V\lp\vecx_0(t)\rp \cdot
 \nabla V\lp\vecx_0(s)\rp|t-s| + \cdots.
 \enqa
It is seen that the phase $X_1$ to first order is simply given by

\beq X_1 = -\frac{v}{4K}\int ds\:\int dt\: \nabla V\lp\vecx_0(t)\rp \cdot
 \nabla V\lp\vecx_0(s)\rp|t-s| + \cdots,
\enq
 and that the second cumulant contains only terms of order $1/K^2$ and higher.
 As we have just argued, the difference between $K$ and $k$ plays here no role. Therefore, the leading
 phase $\chi_0(\vecb)$ and the first order phase $\chi_1(\vecb)$ in the high-$k$ expansion 
 are given by
 \beq
 \boxed{\chi_0(\vecb) = -\int dt\:V\lp \vecb + \frac \vecK m t\rp}
 \enq
 
 \noindent and
 
\beq\label{tau1}\boxed{\chi_1(\vecb) = \frac v {4K} \int dt\:\int ds\:\nabla V\lp \vecb + \frac \vecK m t\rp\cdot\nabla V\lp \vecb + \frac \vecK m s\rp|t-s|.}\enq

\noindent For spherically symmetric potentials, $\chi_0(\vecb)$ can be immediately rewritten as
\beq\chi_0(b) = -\frac 1 v \int dZ\:V\lp r\rp,\enq
while we show in Appendix \ref{reductiontau1} that 
\beq \chi_1(b) = -\frac 1{v^2K} \lp 1 + b\frac{d}{db}\rp\int_0^{\infty}dZ\: V^2(r).\enq 
In both of these expressions, the argument $r$ of the potential is meant to be $\sqrt{b^2 + Z^2}$.
These two phases coincide exactly with the first two terms of the eikonal expansion from Wallace. Thus the first cumulant of the variational procedure contains both the leading order as well as the first order of the high energy expansion. As we will see, it contains even the imaginary part of the second order as well!

\paragraph{Second Order.}
The second order term is a bit more subtle. There are three sorts of contributions to it.\newline
First, the two cumulants $\lambda_1$ and $\lambda_2$ contain terms of that order. While $\lambda_1$ contains only a complex part, that we will denote $\tautwo(\vecb)$, the second cumulant contains to that order only a real part, which we call $\omtwovar(\vecb)$. Indeed, the expression (\ref{lambda2}) we derived for the second cumulant can be rewritten as
\beq \lambda_2 = \sum_{|\alpha| \geq 2}\frac{1}{\alpha !}\left(\frac {-iv} {2K}\right)^{|\alpha|}\int dt_1\int dt_2\: D^\alpha V(\vecx(t_1)) D^\alpha V(\vecx (t_2))\:|t_1-t_2|^{|\alpha|},\enq
with $v = K/m$. The first term of this sum, with the straight line trajectory $\vecx_0(t)$ as argument in the potential functions, is already second order in $1/K$, and thus also in $1/k$, and is purely real. All other terms are of higher order.\newline
Second, the fact that the factor $iK/m$ stands in front of the T-matrix, and not $ik/m$, gives rise to a second order contribution. We will denote the imaginary and real contributions with $\tauk(\vecb)$ and $\omk(\vecb)$ respectively.\newline
Third, the term of order zero in our eikonal representation was found to be (see equation (\ref{chi_0})), after the substitution $Z = t K/m $,
\beq
 -\frac m K \int dZ\: V(\vecb + Z\hat \vecK ) = -\frac m k \lp 1-\frac{q^2}{4k^2}+\cdots\rp^{-1/2}\int dZ\: V(\vecb + Z\hat \vecK ).
\enq
Although we could ignore the difference in first order, it must be taken into account in second order. We will call the imaginary contribution $\omk(\vecb)$ and the real one $\omkk(\vecb)$.
\noindent\paragraph{Computation of $\tautwo$.} By definition this is done by expanding the complex quantities $X_0(\vecb)$ and $X_1(\vecb)$ up to second order. Let us first consider $X_1$. Since $X_1(\vecb)$ is already first order, we need to expand once around the leading trajectory $\vecx_0(t)$.
\beqa
X_1 &=& -\frac 1 {4m}\int dt\int ds\:\gradt\cdot\grads |t-s| \nonumber\\\nonumber
&\approx&  -\frac 1 {4m}\int dt\int ds\:\nabla V(\vecx_0(t))\cdot\nabla V(\vecx_0(s)) |t-s| \nonumber\\
&& + 2\frac{1}{8m^2}\int dt\:ds\:du\nabla_i\nabla_j V(\vecx_0(t))\nabla_j V(\vecx_0(s))\nabla_i V(\vecx_0(u))|t-s||t-u|. \nonumber
\enqa
Similarly, one has to expand $X_0$ to get the second order contribution. Starting from
\beq 
X_0 = -\int dt\: V(\vecx(t)),\enq
one gets, after expanding around $\vecx_0(t)$ to second order,
\beqa
&&X_0 \approx -\int dt\:V(\vecx_0(t)) + \frac 1 {2m} \int dt\:\int ds\:\nabla V(\vecx_0(t))\cdot\nabla V(\vecx(s))|t-s|\nonumber\\
 && -\frac{1}{2}\left(\frac {-1}{2m}\right)^2\int dt\:ds\:du \nabla_i\nabla_jV(\vecx_0(t))\nabla_iV(\vecx(s))\nabla_jV(\vecx(u))|t-s||t-u|\nonumber.\enqa
The second order term is then given by the third term in the last equality with $\vecx_0(s)$ and $\vecx_0(u)$ instead of
$\vecx(s)$ and $\vecx(u)$, while the second term must be again expanded around $\vecx_0(s)$ and gives rise to another contribution of the very same form. After getting the numerical factors right, the total contribution from $X_0$ and $X_1$ is
\beq
\boxed{\tautwo=-\frac{v^2}{8K^2}\int dt\:ds\:du\nabla_i\nabla_j V(\vecx_0(t))\nabla_j V(\vecx_0(s))\nabla_i V(\vecx_0(u))|t-s||t-u|,}
\enq
with $v = K/m$. For spherically symmetric potentials, it is shown in the appendix \ref{tau2} that it reduces to the form
\beqa \tautwo(b) = \nonumber &-&\frac{1}{v^3K^2}\left(1 +\frac53b\frac{d}{db}+\frac13b^2\frac{d^2}{db^2}\right)\int_0^{\infty}dZ\:V^3(r) - \\
&-&\frac{1}{8K^2}\left[\chi_0(\chi_0^{'})^2+\frac13b(\chi_0^{'})^3\right].\enqa

\paragraph{Computation of $\tauk$ and $\omk$.}
These two terms are the complex and real contributions coming from the $K$ factor in front of the T-matrix:
\beqa
T_{i\rightarrow f}^{(3-3)}\nonumber &=& i\frac{K}{m} \int d^2b\:e^{-i\vecq\cdot\vecb}\lp \exp\lp i\lambda_1-\frac 12 \lambda_2\rp-1\rp \\
&=& i\frac{k}{m} \int d^2b\:\lp1-\frac{q^2}{8k^2} - \nonumber \cdots \rp e^{-i\vecq\cdot\vecb}\lp \exp\left[ i\lambda_1-\frac 12 \lambda_2\right]-1\rp.
\enqa 
To absorb this $q^2$ term into a correction phase depending on $\vecb$ only, we observe that
\beq
\left(1-\frac{q^2}{4k^2}\right)\exp\lp -i\vecq\cdot\vecb \rp = \lp1+\frac{\Delta_b}{4k^2}\rp \exp\lp -i\vecq\cdot\vecb \rp,
\enq
with $\Delta_b$ being the 2-dimensional Laplacian operator with respect to $\vecb$. It is now possible to integrate by parts and to incorporate the correction into the phase. In the appendix \ref{2order}, we show that the final results are
\beq
\boxed{\tauk(\vecb) = \frac{1}{8k^2}\Delta_b\chi_0}
\enq
and
\beq
\boxed{\omk(\vecb)= \frac{1}{8k^2} (\nabla_b\chi_0)^2.}
\enq
\noindent\paragraph{Computation of $\taukk$ and $\omkk$.} As explained above, these two terms appears because of the factor of $K$ present in the leading order phase in the eikonal representation. As for $\tauk$ and $\omk$, they come with a factor $q^2$ that can be transformed away by a Laplacian operator acting on the scattering phases after an integration by parts. Detailed calculations are provided in appendix \ref{2order}. The results are
\beq
 \boxed{\taukk(\vecb) = \frac{1}{8k^2}\lp (\nabla_b\chi_0)^2\chi_0 - \Delta_b\chi_0 \rp,}
\enq
and
\beq
 \boxed{\omkk(\vecb)= -\frac{1}{8k^2}\lp 2\:\lp  \nabla_b\chi_0\rp^2+\chi_0\Delta_b\chi_0  \rp,}
\enq
\paragraph{Computation of $\omtwovar$.} We already mentioned that $\omtwovar$ is the first term in the infinite sum (\ref{lambda2}) that describes the second cumulant. Getting all the numerical factors right, it thus reads
\beq
\boxed{\omtwovar(\vecb) = -\frac{v^2}{16K^2}\int dt\:\int ds\:  |t-s|^2\:\nabla_i\nabla_j V\lp\vecx_0(t)\rp \nabla_i\nabla_jV\lp\vecx_0(s)\rp}
\enq
%
%
Appendix B2 provides the derivation of its reduced form for a spherically symmetric potential. There it is shown that
\beq \omtwovar(b) = \frac{1}{8K^2}\left[2\chi_0^{'2}+\chi_0^{''}\left(\chi_0+b\chi_0^{'}\right)+\frac 1 b \chi_0 \chi_0^{'}\right].
\enq  
\newline
It suffices now to sum over all the contributions to obtain our second order scattering phases. We have, written here explicitly only for potentials with spherical symmetry,
\beqa
\omega_2(b) &=& \omtwovar(b) + \omk(b) + \omk(b)\nonumber\\
&=&  \frac{b}{8k^2}\chi_0'\Delta_b\chi_0,
\enqa
with $\Delta_b\chi_0 = \chi_0^{''} + \chi_0'/b $. For the imaginary term,
\beqa
\chi_2(b) &=& \tautwo(b) + \tauk(b) + \taukk(b)\nonumber\\
&=& \boxed{ -\frac{1}{v^3k^2}\left(1 +\frac 53 b\frac{d}{db} + \frac{1}{3}b^2\frac{d^2}{db^2}\right)\int_0^{\infty}dZ\: V^3(r) -
 \frac{b\chi_0^{'3}}{24k^2}.}
\enqa
These expressions turn out to be exactly the same as those in the expansion from Wallace (see expressions (\ref{x0w}) to (\ref{omega2w})). It is remarkable that the variational procedure, in the very first cumulant, contains already all imaginary terms up to the second order, plus a part from the real term $\omega_2$. The first correction to the variational procedure, i.e. the second cumulant, then completes it.
\subsection{Unitarity of the Approximation}
The investigation of the unitarity of the scattering matrix, or the validity of the optical theorem, in our variational approximation is not an easy task. Even at the very beginning, although (\ref{T33}) and (\ref{T31}) are exact representations, its validity is but self-evident. We have seen that the variational approximation is close to an impact parameter representation of the T-matrix. Especially, we just showed that it reduces to the eikonal expansion for high energies. We will try to gain some insights in this question within a very brief look at the unitary of impact parameters representations. We will consider in the following central potentials only.\newline
We denote by $\chi$ the imaginary and by $\omega$ the real part of the scattering phases, and by $S(b)$ the quantity
\beq
S(b):=e^{i\chi(b)-\omega(b)},
\enq
so that the T-matrix elements in the impact parameter (eikonal) formalism are given by (see equations (\ref{T_I}) and (\ref{T_II}))
\beq \label{Teik}
T_{i\rightarrow f}= i\frac k m \int d^2b \:e^{-i\vecq\cdot\vecb}\:T(b),\quad T(b) = S(b) -1.
\enq
The optical theorem reads 
\beq
\sigma_{tot} = \frac{4\pi}{k} \Im f(0).
\enq
It relates the total cross section to the imaginary part of the scattering amplitude $f$, and follows directly from the condition of unitarity of the scattering matrix (a derivation can be found, for example, in \cite{Messiah}). The scattering amplitude is in turn related to the T-matrix through
\beq
 f(\Omega)= -\frac{m}{2\pi}T_{i\rightarrow f}.\quad (\hbar = 1)
\enq
When $T$ is written in the form (\ref{Teik}), the following relation must therefore hold: 
\beq
\sigma_{tot} = -2\int d^2b\: \left[ \Re(S) -1\right].
\enq
The total cross section is, on the other hand, by definition
\beqa
\sigma_{tot} &=& \int d\Omega \left|f\right|^2\\
&=& \frac{1}{(2\pi)^2}\int  \frac {d\Omega}{k^2} \int d^2b\int d^2b'\:e^{-i\vecq\vecb}e^{i\vecq\vecb'}T(b)T^{*}(b').
\enqa 
Since $q= 2k\sin(\theta/2)$, it holds that
\beq
q\:dq = 2k^2\sin\left(\frac \theta2\right)\cos\left(\frac\theta 2\right)d\theta = k^2\sin(\theta)d\theta,
\enq
with implies in turn
\beq
\int\frac {d\Omega(\theta,\phi)}{k^2} = \int_{q \leq 2k} d^2q.
\enq
The range of integration is restricted to the physically possible momentum transfers whose value does not exceed $2k$. However, one can also integrate over all possible values of $\vecq$, and then subtract the non-physical scattering cross section, for which $q \ge 2k$:
\beq
\sigma_{tot}  = \frac{1}{(2\pi)^2}\int d^2 q\int d^2b\int d^2b'\:e^{-i\vecq\cdot\vecb}e^{i\vecq\cdot\vecb'}T(b)T^{*}(b') -\sigma_{\mathrm{np}} = \int d^2b \:\left|T(b)\right|^2 -\sigma_{\mathrm{np}},
\enq
where $\sigma_{np}$ is
\beq
\sigma_{\mathrm{np}} := \frac{1}{(2\pi)^2}\int_{q \geq 2k} d^2 q\left|\int d^2b\:e^{-i\vecq\cdot\vecb}T(b)\right|^2\geq 0
\enq
If the optical theorem is to hold, the following relation must thus be fulfilled:
\beqa
\sigma_{\mathrm{np}} &\stackrel{!}{=}& \int d^2b \:\left\{\left|T(b)\right|^2 + 2\left[\Re(S(b)) - 1\right]\right\} \\
&=& \int d^2b\:\left[\left|S(b)\right|^2 -1\right]\\
&=&\int d^2b\:\left[e^{-2\omega(b)} - 1\right].
\enqa
Since $\sigma_{\mathrm{np}}$ is positive definite, the real part $\omega$ of the scattering phases is thus seen to play an essential role in the theory. At high energy, we have seen that the real term $\omega$ predicted by the variational approximation is the term(\ref{omega2w}),
\beq
\omega_2(b) = \frac{1}{16k^2}\frac1b\frac d{db}\left(b\chi_0'\right)^2.
\enq
The non-physical cross section can then be calculated as
\beqa
\sigma_{\mathrm{np}} \stackrel{k\rightarrow\infty}{\rightarrow} -2\int d^2b \:\omega_2(b) &=& -\frac{4\pi}{16k^2}\int_0^{\infty}db\:\frac {d}{db}\left(b\chi_0'\right)^2\\
&=& \frac{\pi}{4k^2}\lim_{b\rightarrow0}\left(b\chi_0'\right)^2.
\enqa
This expression vanishes for potentials that are regular at the origin, but this contribution is finite, for instance, for a Yukawa potential. The question how well unitarity is fulfilled at high energy thus depends strongly on the specific potential.\newline
One of the main differences between the variational approximation in the eikonal representation and in the ray representation is, as we will see, that in the latter, such a real part is contained in the approximation, while we have seen that in the former it appears in the second cumulant only. It should therefore not be considered as a surprise that the variational approximation in the ray representation will turn out to be superior than its eikonal counterpart.


%% file: RayRepresentation.tex
\newpage
\section{Variational Approximation in the Ray Representation} \label{RayRepresentation}
Besides the facts that the ray representation does not treat the parallel and perpendicular components on an equal footing, and that the trial function $\C(t)$ is yet only a scalar, all calculations in the ray representation resemble very much those we have performed in the eikonal representation. However, we will encounter here new interesting features, that render this approach very different from the eikonal one of the last section. We will proceed in the very same order, to obtain first, the approximation
\beq
T^{(3-1)}_{i\rightarrow f} = i\frac K m \int d^2b\:e^{-i\vecb\cdot\vecq}\:T_1^{(3-1)}(\vecb,\vecq,\vecK),
\enq
that contains only the first cumulant, and, second,
\beq
T^{(3-1)}_{i\rightarrow f} = i\frac K m \int d^2b\:e^{-i\vecb\cdot\vecq}\:T_2^{(3-1)}(\vecb,\vecq,\vecK),
\enq
with the first correction (second cumulant) included.
\paragraph{First Cumulant.}
Obviously, the first cumulant is given by the suitably adapted formula (\ref{l133}), i.e.
\beq \label{l1ray}
\cumulr1 = \la\chi_{ray}\rat+\frac 1 {2m}\left[\left(\vecB,\vecB\right)-\left(\C,\C\right)\right],
\enq
and the variational equations for $\vecB$ and $\C$ are the equivalent of (\ref{varB}) and (\ref{varC}):
\beq \label{varBr}\vecB(t) = -m\:\frac{\delta\la\chi_{ray}\rat(\vecB,\C)}{\delta\vecB(t)},\enq
respectively
\beq \label{varCr}\C(t) = m\:\frac{\delta\la\chi_{ray}\rat(\vecB,\C)}{\delta\C(t)}.\enq
\paragraph{Computation of $\la\chi_{ray}\rat$.}
Since the computations are very similar to those we already have done, they are only reported in appendix \ref{chik}. There it is shown that the expectation value
$\la\chi_{ray}\rat$ has the following form:
\beqa
&&\la\chi_{ray}\rat_{ray} \nonumber = -\int dt\:\int \frac{d\vecp}{(2\pi)^3}\widetilde V (\vecp)\exp\lp i\vecp\left\{\vecb+\frac{\vecp_{ray}(t)}{m}t\right\}\rp\exp\lp-\frac i{2m}|t|\vecp_{\perp}^2\rp\\
&&\cdot\exp\lp-\frac{i}{2m}\vecp\left\{\int ds\:\vecB(s)\left[\sgn(t-s)-\sgn_{\perp}(-s)\right]+\C(s)\sgn_\parallel(-s)\right\}\rp.
\enqa
The symbols $\perp$ (or $\parallel$), such as in $\sgn_\perp(s)$, have the meaning that it multiplies only the components of the 3-dimensional vector that is involved in the product that are perpendicular (or respectively parallel) to $\vecK$  (for example $\vecp\:\sgn_\perp(s) = \vecp_\perp\sgn(s)$). We will use this convention in the following, as it permits to render most expressions more concisely and elegantly.\newline
This expression for $\la\chi_{ray}\rat$ contains something new with respect to the (3-3) version. The exponential factor that contains a quadratic term in $\vecp_\perp$ does not permit to return immediately to real space, as we could do in the eikonal representation. Nevertheless, we interpret the following quantity $V_{\sigma(t)}$ as being a new, effective potential, defined in Fourier space through the (Fresnel) Gaussian transformation
\beq \label{Veff}
\boxed{\widetilde V_{\sigma(t)}(\vecp) := \widetilde V(\vecp)\:\exp\lp -\frac12\sigma(t)\vecp_\perp^2 \rp, \quad \mathrm{with} \quad \sigma(t) = \frac{i}{m}|t|.}
\enq
This new potential, in real space, is the convolution of the potential $V$ with a Gaussian filtering function with typical filter size $\sqrt\sigma$. However, the filter being complex, this effective potential is \textit{a priori} also a complex quantity. This is an essential feature of the ray representation. Note that at time $t = 0$, the effective potential is the same as the potential $V$, and that is also vanishes in the limit of asymptotic times.\newline The arguments that multiply the linear term in $\vecp$ can be then defined as the trajectory
\beq
\vecz(t) = \vecb + \frac {\vecp_{ray}(t)}{m}t -\frac{i}{2m}\left[\int ds\:\vecB(s)\left[\sgn(t-s)+\sgn_{\perp}(s)\right]-\hat\vecK\C(s)\sgn(s)\right],
\enq
and the final result for $\la\chi_{ray}\rat$ has a similar form as in the eikonal representation  
  
\beq\label{chiray}\la\chi_{ray}\rat = -\int dt\:V_{\sigma(t)}(\vecz(t)).\enq
\subsection{Variational Equations and the Variational Trajectory}
\begin{figure}[htbp]
	\centering
		\includegraphics[width=11cm]{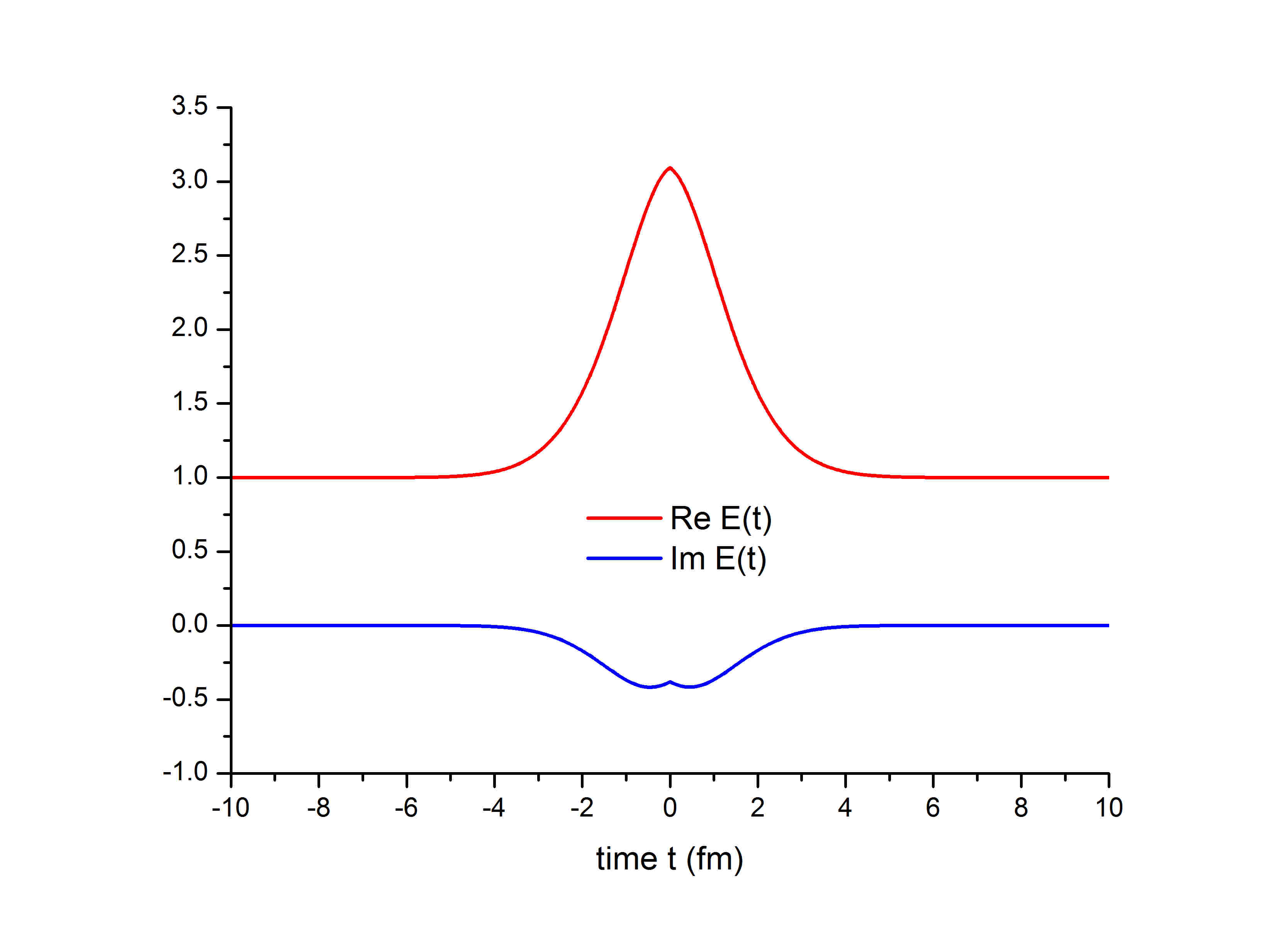}
		\caption {The real (red line) and imaginary part (blue line) of the "energy" of the variational trajectory in the ray representation, as a function of time, in units of the scattering energy $\frac {k^2}{2m}$, for a particular choice of parameters and a Gaussian potential. With the parameters chosen, the target is about 2 fm large, and the incoming velocity about $0.5c$. This implies a typical interaction time scale, expressed in terms of length, of 4 fm. }
	\label{fig::Energy}
\end{figure}

We can now proceed to solve the variational equations in the ray representation. Just as in the (3-3) case, setting $\vecB$ and $\C$ to zero reduces the trajectory to the straight lines $\vecb + t\:\vecp_{ray}(t)/m $. However, if these are taken to satisfy the variational equations (\ref{varBr}) and (\ref{varCr}), it must first hold that

\beq \label{Bray}
\vecB(t) = -\frac12 \int ds\: \nabla \Vs s(\vecz(s))\left[ \sgn(s-t) -\sgn_\parallel (-t)\right],
\enq
and
\beq \label{Cray}
\C(t)   = \frac12 \int ds\:\nabla\Vs s(\vecz(s))\sgn_\parallel(-t).
\enq
and, second, after inserting these expressions into the trajectory $\vecz$, we get
\beq \label{zt}\boxed{\vecz(t) = \vecb +\frac\vecK m t +\frac{\vecq}{2m}|t| -\frac1{2m}\int ds\:\nabla \Vs s(\vecz(s))\left[ |t-s| -|t|_\perp -|s|_{\perp} \right].}
\enq
This trajectory is a complex trajectory, and possesses some properties analogous to those in the eikonal representation.
The component of the trajectory that is parallel to $\vecK$ also behaves classically. We have, by differentiating twice,
\beq \label{ztpar}
\ddot\vecz_\parallel(t) = -\frac 1 m \nabla_\parallel \Vs t(\vecz(t)).
\enq 
However, the perpendicular part gets a kick at $t = 0$. Indeed, the two terms that contain $|t|$ give rise to $2\delta(t)$, and we get
\beq \label{ztperp}
\ddot\vecz_\perp(t) = -\frac 1 m \nabla_\perp \Vs t(\vecz(t)) + \delta(t)\left( \frac{\vecq}{m} +\frac1m \int ds\:\nabla_\perp \Vs s(\vecz(s)) \right).
\enq 
\paragraph{Asymptotic Behavior, Constants of Motion: Symmetric potentials.}
The effective potential $\Vs t(\vecz) $ has a special dependence on the perpendicular components of $\vecz$. However, it still depends only on the modulus of its parallel and perpendicular components, if $V$ does so, although in different ways. In that case, by the very same argument as in the eikonal representation, the symmetry properties
\beq \label{symm}
\vecz_\perp(-t) = \vecz_\perp(t),\quad \vecz_\parallel(-t) = -\vecz_\parallel(t)
\enq still hold, although $\Vs t$ is generally not spherically symmetric. The defining equation (\ref{zt}) can be then rewritten such that the unnatural subscripts are eliminated,
\beq \label{zzt}\boxed{\vecz(t) = \vecb +\frac\vecK m t +\frac{\vecq}{2m}|t| -\frac1{2m}\int ds\:\nabla \Vs s(\vecz(s))\left[ |t-s| -|t| -|s| \right].}
\enq
The trajectory (\ref{zzt}) exhibits interesting properties. First, for very large times, it holds that
\beq
|t-s|-|t|-|s| \stackrel{t\rightarrow\pm\infty}{\longrightarrow} \mp s -|s|,\enq so that the integral part of the trajectory goes to a constant. We name its parallel component $\vecB_{ray}/2$, and its perpendicular $\vecC_{ray}$, 
\beqa\label{ray} \vecB_{ray}  &=& \frac1m\int ds\:\nabla\Vs s (\vecz(s))s
\\ \vecC_{ray}  &=&\frac1{2m}\int ds\:\nabla\Vs s (\vecz(s))|s|
\enqa
so that
\beq \label{asymp}\vecz(t) \stackrel{t\rightarrow\pm\infty}{\longrightarrow}\vecb +\vecC_{ray} \pm \frac12 \vecB_{ray} +\frac\vecK m t \pm \frac{\vecq}{2m}t :=\vecb_{f,i}^{ray} + t\:\vecv_{f,i}^{ray}.
\enq
This implies immediately
\beq
\dot\vecz(t) \stackrel{t\rightarrow\pm\infty}{\longrightarrow} \frac\vecK m  \pm \frac\vecq{2m} =  \left\{\begin{array}{rrr}
\veck_f/m & t\rightarrow +\infty \\ \veck_i/m & t\rightarrow -\infty \end{array} \right. 
\enq
The asymptotic momenta of the variational trajectory (\ref{zzt}) are thus precisely those of the scattering process, $\veck_i$ and $\veck_f$. Both the mean momentum $\vecK$ and the momentum transfer $\vecq$ of the scattering process have now the same meaning in the variational trajectory.\newline Furthermore, the "energy", which is now a complex number,
\beq E(t) = \frac 12 m \dot \vecz ^2(t) + \Vs t(\vecz(t)) \enq
is at infinity
\beq
E (\pm\infty) = \frac {\veck_i^2}{2m} = \frac{\veck_f^2}{2m},\label{ErRay}
\enq
since the potential vanishes. This is precisely the scattering energy of the process we are investigating.
However, there is no conservation of energy for finite times, since the effective potential has an explicit time dependence.
Nevertheless, the kick at time $t = 0$ is elastic. This can be seen from the fact at $t = 0$ the perpendicular components of $\dot\vecz(t)$ get reversed, while the longitudinal part stays untouched: from (\ref{zzt}) and (\ref{symm}) it follows
\beqa
\dot\vecz(0_\pm) = \frac\vecK m \pm\frac\vecq{2m} -\frac1{2m}\int ds\:\nabla_\parallel\Vs s\sgn(-s)\mp\frac{1}{2m}\int ds\:\nabla_\perp\Vs s,\enqa
which is to say
\beq \label{zpointzero}
\dot\vecz_\parallel(0_+) = \dot\vecz_\parallel(0_-),\quad \mathrm{and}\quad \dot \vecz_\perp(0_+) = -\dot\vecz_\perp(0_-).
\enq
Also, it holds
\beq \label{zero}
\vecz(0) = \vecb.
\enq
A particle following this trajectory thus starts with the right energy of the scattering process, enters the interaction region where its energy changes, and becomes complex (see figure \ref{fig::Energy} for an illustration), at time $t = 0$ experiences an elastic reflection on the plane perpendicular to $\vecK$, and then gets out with the same energy as when it entered.
\newline

%
\noindent The analysis of the angular momentum for spherically symmetric potentials also provides interesting insights. Since the effective potential itself does not possess rotational symmetry, the conservation of the three components of the angular momentum is not achieved. However, since $\Vs t$ is still symmetrical in the perpendicular plane, the conservation of its longitudinal component is ensured, except at $t = 0$, where the kick takes place, when it suffers a change. This is seen from its definition
\beq \label{Lpar}
\vecL_\parallel = m\vecz_\perp(t)\wedge\dot\vecz_\perp(t),
\enq 
together with the fact (see conditions \ref{cond}) that
\beq
\nabla_\perp\Vs t(\vecz(t)) = f(\vecz_\perp^2(t),\vecz_\parallel^2(t))\:\vecz_\perp(t),
\enq
for a scalar function $f$, from which follows, for any non-zero time $t$,
\beq
\frac d{dt}\vecL_\parallel(t) = m\vecz_\perp(t)\wedge\ddot\vecz_\perp(t) = -f\:\vecz_\perp(t)\wedge\vecz_\perp(t) =0 \quad(t \neq 0).
\enq
The other components of $\vecL$ would mix in this last equality the functions of $f$ and $g$ of (\ref{cond}), so that they are in general not conserved. Let us have a look at the change in $\vecL_\parallel$ at time $t=0$: From (\ref{zpointzero}) and (\ref{zero}) it follows at once
\beq
\vecL_\parallel(t = 0_{\pm}) = m\:\vecz_\perp(0)\wedge\dot\vecz_\perp(0_\pm) = \pm m\:\vecb\wedge\dot\vecz_\perp(0_+),\enq
so that $\vecL_\parallel$ just flips sign at $0$. The asymptotic angular momentum can be calculated from (\ref{asymp}) and (\ref{Lpar}). It reads
\beq
\vecL_\parallel(\pm\infty) = m \left(\vecb + \vecC_{ray}\right)\wedge\pm \frac\vecq{2m}.
\enq
While the asymptotic energy is the same as in the reference straight line trajectory in the ray representation, $\vecb + t\frac{\vecp_{ray}(t)}{m}$, and is thus unaffected by the variational procedure, the angular momentum gets a correction in terms of the quantity $\vecC_{ray}$.
\subsection{The Scattering Phases}
The scattering phases in the ray representation are obtained from (\ref{l1ray}). We already computed the first phase $\la\chi_{ray}\rat$ with (\ref{chiray}) as a result. The second phase is found by inserting the variational equations (\ref{Bray}) and (\ref{Cray}) into the second part of (\ref{l1ray}). It becomes

\beq T_1^{(3-1)}(\vecb,\vecq,\vecK) = \exp  \left(iX_0^{(3-1)} + iX_1^{(3-1)}\right) -1\enq
with the definitions:  

\beq \boxed{X_0^{(3-1)}(\vecb,\vecq,\vecK) =  -\int dt\: \Vs t(\vecz(t)) = \la\chi_{ray}\rat,}\enq
and
\beq \boxed{X_1^{(3-1)}(\vecb,\vecq,\vecK) = -\frac 1 {4m}\int dt\int ds\:\nabla\Vs t (\vecz(t))\nabla\Vs s (\vecz(s))\left[|t-s| -|t|_\perp -|s|_\perp\right].}\enq
Again, if the potential $V$ satifies the conditions (\ref{cond}), the subscripts $\perp$ can be left out. The main difference between these two terms and their eikonal counterparts is that they are both complex quantities, and contain thus much more information in comparison. 
In the case of a spherically symmetric potential, it holds
\beq T_1^{(3-1)}(\vecb,\vecq,\vecK) = T_1^{(3-1)}(b,q,K,\vecb\cdot\vecq),\enq
since $\vecq$ and $\vecb$ are perpendicular to $\vecK$.
The relation (\ref{rel}) between these two phases has in this case the form:
\beq
X_0^{(3-1)} -X_1^{(3-1)} = \vecK\cdot\vecB_{ray} + \vecq\cdot\vecC_{ray} + \int dt\:\left[\frac{k^2}{2m}-E(t)\right], \quad k = |\veck_i| = |\veck_f|.
\enq 
The integral on the right is well defined, since, as we have shown above, the integrand takes on non-zero values only near the scattering region (see equation (\ref{ErRay}), or figure \ref{fig::Energy}).
The derivation is in all aspects similar. After inspection of (\ref{zt}), (\ref{ztpar}) and (\ref{ztperp}), the second phase can namely written as
\beqa
X_1^{(3-1)} &=&\frac 12\int dt\:\nabla\Vs t (\vecz(t))\left[\vecz(t)-\vecb-\frac \vecK m t -\frac\vecq{2m}|t|\right]\nonumber  \\ 
&=& -\frac m2\int dt\:\ddot \vecz(t)\left[\vecz(t)-\vecb-\frac \vecK m t -\frac\vecq{2m}|t|  \right]\nonumber \\
&&+\frac m 2 \left[\frac\vecq m +\frac 1m \int ds\:\nabla_\perp\Vs s\right]\left[\vecz(0)-\vecb\right]
\enqa
The last term is zero, so that we get
\beq X_1^{(3-1)} = -\frac m2\int dt\:\ddot \vecz(t)\left[\vecz(t)-\vecb-\frac \vecK m t -\frac\vecq{2m}|t|  \right].\enq
We use the same trick as we already did to decompose the integral in different terms,
\beqa
\left.X_1^{(3-1)}\right|_T &=& \left.\frac m2\dot\vecz(t)\left[\vecz(t)-\vecb-\frac \vecK m t -\frac\vecq{2m}|t|  \right]\right|_{t=-T}^{t=+T}\\
&&+\frac m 2 \int dt\:\dot\vecz^2(t) -\left.\frac\vecK 2 \vecz(t)\right|_{t=-T}^{t=+T}-\frac\vecq{4}\int_{-T}^{+T}dt\:\dot\vecz(t)\sgn(t),
\enqa
and recover $X_0^{(3-1)}$, since
\beqa
\frac m 2 \int_{-T}^{+T} dt\:\dot\vecz^2(t) &=&\int_{-T}^{+T}dt\:\left[E(t) -\Vs t(\vecz(t))\right]\nonumber\\
 &=& \int_{-T}^{+T}dt\:E(t) + \left.X_0^{(3-1)}\right|_{T}.
\enqa
After inserting the asymptotic behaviour (\ref{asymp}) of the trajectory in the other terms and some algebra, one obtains what was claimed.

\subsection{First Correction to the Variational Approximation}
The first correction to the variational procedure is given by the second cumulant $\lambda_2^{(3-1)}$. The computations are essentially a copy-paste of those done in the eikonal representation. Especially, formula (\ref{l1var}) still holds,
\beq
\lambda_2^{(3-1)} = \la\chi_{ray}^2\rat -\la\chi_{ray}\rat^2  -2iX_1^{(3-1)}. \label{l231}
\enq 
The evaluation of $\la\chi_{ray}^n\rat$ is done in appendix \ref{chik}, one has, for $n =2$,
\beqa
&&\chir 2 =\nonumber\int dt_1\int dt_2\:\int\prodi i12 \frac{d\vecp_i}{(2\pi)^3}\widetilde V(\vecp_i)\exp\left(i\sumi i12 \vecp_i\cdot\vecz(t_i)\right)\cdot\\
&\cdot&\exp\lp\frac{i}{2m}\vecp_1\cdot\vecp_2 \big[|t_1-t_2|-|t_1|_\perp-|t_2|_\perp\big]\rp\cdot \exp\left(-\frac i{2m}\left[\vecp_{1\perp}^2|t_1|+\vecp_{2\perp}^2|t_2|\right]\right)\nonumber.
\enqa
With the definition (\ref{Veff}) of the effective potential it reads
\beqa
\chir 2 &=&\nonumber\int dt_1\int dt_2\:\int\prodi i12 \frac{d\vecp_i}{(2\pi)^3}\widetilde V_{\sigma(t_i)}(\vecp_i)\exp\left(i\sumi i12 \vecp_i\cdot\vecz(t_i)\right)\cdot\\
&\cdot&\exp\lp\frac{i}{2m}\vecp_1\cdot\vecp_2 \big[|t_1-t_2|-|t_1|_\perp-|t_2|_\perp\big]\rp.
\enqa
The only differences with respect to the eikonal representation are thus the presence of the effective potential and the terms $|t_i|_\perp$ in the exponent. One sees that the first two terms in the developpment of this exponential give rise to $X_0^{(3-1)}$ squared and $2iX_1^{(3-3)}$ respectively, so that the second cumulant is that expression, starting from the third term in the expansion of the exponential function. 
\subsection{High Energy Expansion}
As in the eikonal representation, we fix the velocity $k/m$ and expand the trajectory and the corresponding phases in inverse powers of $k$. Since we will not investigate in much detail beyond the first order, we will not make any difference between $k$ and $K$. 
\newline We first have a look to the properties of the effective potential in this approach. From its very definition (\ref{Veff}), one has
\beqa
\widetilde V _{\sigma(t)}  (\vecp) = \widetilde V(\vecp) \exp\lp -i \frac v{2K}|t|\vecp_\perp^2\rp \approx \widetilde V(\vecp) - i\frac{v}{2K}|t|\vecp_\perp^2\widetilde V (\vecp), 
\enqa 
so that for large $K$, or for small times, the effective potential is $V$ with a complex correction,
\beq
\Vs t \approx V + i\frac v{2K}|t|\Delta_\perp V.
\enq
Especially, the imaginary part of the variational trajectory behaves, under such conditions (for example near the scattering region), under the influence not of $V$, but of the (perpendicular) Laplacian of $V$. This imaginary trajectory is clearly a second order effect. Indeed, the zeroth order $\vecz_0(t)$ is the usual straight line
\beq
\vecz_0(t) = \vecb + \frac\vecK m t + \frac \vecq{2m}|t|, 
\enq
while the first order term $\vecz_1(t)$ is readily obtained through
\beq \label{z1}
\vecz_1(t) = -\frac v {2K}\int ds\:\nabla V(\vecz_0(t))\left[|t-s|-|t|_\perp-|s|_\perp\right],
\enq
where the zeroth order term $V$ of $\Vs s$ had to be used. Thus $\vecz_1(t)$ is still real, while in the second order term of $\vecz(t)$ would appear the first order term of $\Vs s $, which is complex.
\newline 
Let us have a look at the scattering phases to first order. The factor $1/m = v/K$ in front of $ X_1^{(3-1)}$ means that, up to higher order terms,
\beq \label{x1r}
X_1^{(3-1)} \approx -\frac{v}{4K} \int ds\:\int dt\:\nabla V (\vecz_0(s))\nabla V  (\vecz_0(t))\left[|t-s|-|t|_\perp-|s|_\perp\right].
\enq 
On the other hand, the phase $X_0^{(3-1)}$ can be written as
\beqa
&&X_0^{(3-1)} = -\int dt\int \frac {d\vecp}{(2\pi)^3}\widetilde V(\vecp)\exp\lp-\frac i{2m}\vecp_\perp^2|t| \rp\exp\left[i\vecp\cdot\vecz(t)\right]\\
&\approx& -\int dt\int \frac {d\vecp}{(2\pi)^3}\widetilde V(\vecp)\lp 1-\frac i{2m}\vecp_\perp^2|t| \rp\exp\left[i\vecp\cdot\vecz_0(t)\right]\lp 1-i\vecp\cdot\vecz_1(t)\rp \nonumber,
\enqa 
and contains two real parts, one of zeroth order, that we will call $\chi_0^{(3-1)}$, and one of first order, which is recognized, using (\ref{z1}), to be
\beqa \nonumber
&&\int dt\int \frac {d\vecp}{(2\pi)^3}i\vecp \:\widetilde V(\vecp) \:\vecz_1(t)\exp\lp i\vecp\cdot\vecz_0(t)\rp \\
&=& \label{x1r2}\frac v{2K}\int ds\:\int dt\:\nabla V (\vecz_0(s))\nabla V  (\vecz_0(t))\left[|t-s|-|t|_\perp-|s|_\perp\right].
\enqa
However, $X_0^{(3-1)}$ contains to first order also an imaginary part $\omega_1^{(3-1)}$, given by
\beqa
\omega_1^{(3-1)} &=& \nonumber
\frac{v}{2K}\int dt\:|t|\int \frac {d\vecp}{(2\pi)^3} \vecp_\perp^2 \:\widetilde V(\vecp)\exp\lp i\vecp\cdot\vecz_0(t)\rp\\
&=& -\frac v{2K}\Delta_b\int dt\:|t|V(\vecz_0(t)), \label{omega1}
\enqa
where $\Delta_b$ is the two dimensional Laplacian operator acting on $\vecb$ in $\vecz_0(t)$. To sum up, one has after adding (\ref{x1r}),(\ref{x1r2}),(\ref{omega1}), up to higher order terms,
\beq
i\lambda_1^{(3-1)} = i\lp\chi_0^{(3-1)} + \chi_1^{(3-1)}\rp - \omega_1^{(3-1)},
\enq
with
\beq
\boxed{\chi_0^{(3-1)} = -\int dt\:V(\vecz_0(t)),}
\enq
\beq
\boxed{\omega_1^{(3-1)} =  -\frac v{2K}\Delta_b\int dt\:|t|V(\vecz_0(t)),}
\enq
and
\beq
\boxed{\chi_1^{(3-1)} = \frac v{4K}\int ds\:\int dt\:\nabla V (\vecz_0(s))\cdot\nabla V  (\vecz_0(t))\left[|t-s|-|t|_\perp-|s|_\perp\right].}
\enq
These three phases are identical with those obtained in the high-energy limit in \cite{Rosenfelder}. We refer to it for a more complete discussion.

%% file: Numerics.tex
\newpage
\section{Numerical Implementation and Results}\label{numerics}
In this section we test numerically the approximations of the T-Matrix we obtained in this work. More concretely, we compute numerically the differential cross section
\beq
\sigma(\Omega) = \left|f(\Omega)\right|^2 = \frac {m^2}{(2\pi)^2}\left|T_{i\rightarrow f}\right|^2,
\enq 
where $T_{i\rightarrow f}$ are the expressions (\ref{Tap33}) and (\ref{Tap31}). \newline\noindent These are tested against the eikonal expansion of Arbabanel and Itzykson, and partial wave calculations. A sufficient number of partial waves were considered, so that we could consider them as the exact results. 
\subsection{Eikonal Representation}
\begin{figure}[htbp]
	\centering
		\includegraphics[width=10cm]{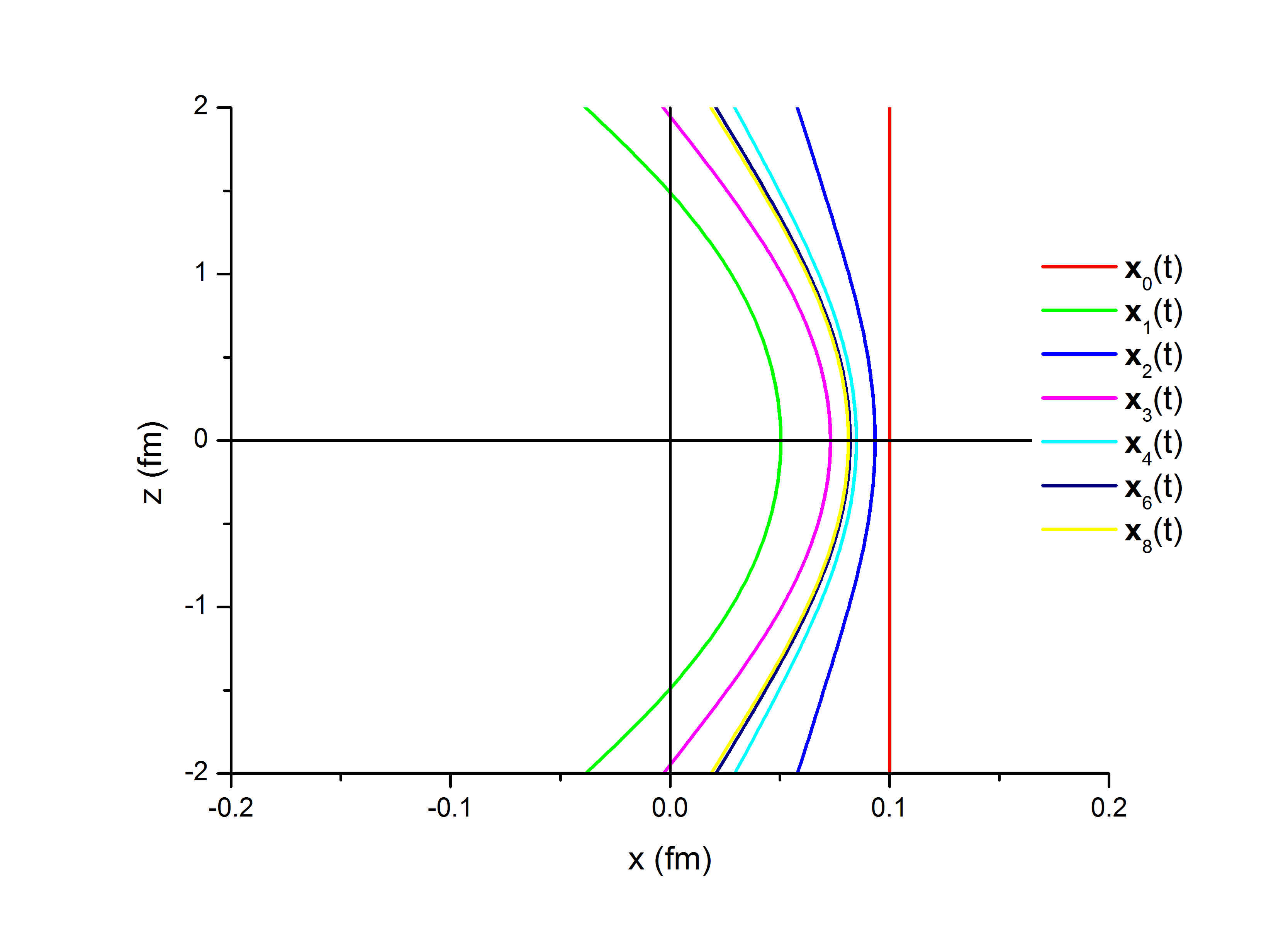}
		\caption {The successive approximations $\vecx_n(t)$ of the variational trajectory $\vecx(t)$, for a particular value of $b = 0.1$ fm, and $K = 2$ fm$^{-1}$. The potential is Gaussian, with the parameters given in the text. The z-scale has been largely stretched, so that the differences in the asymptotic momenta can be clearly seen.}
	\label{fig::GTraj}
\end{figure}

\begin{figure}[htbp]
	\centering
		\includegraphics[width=15cm]{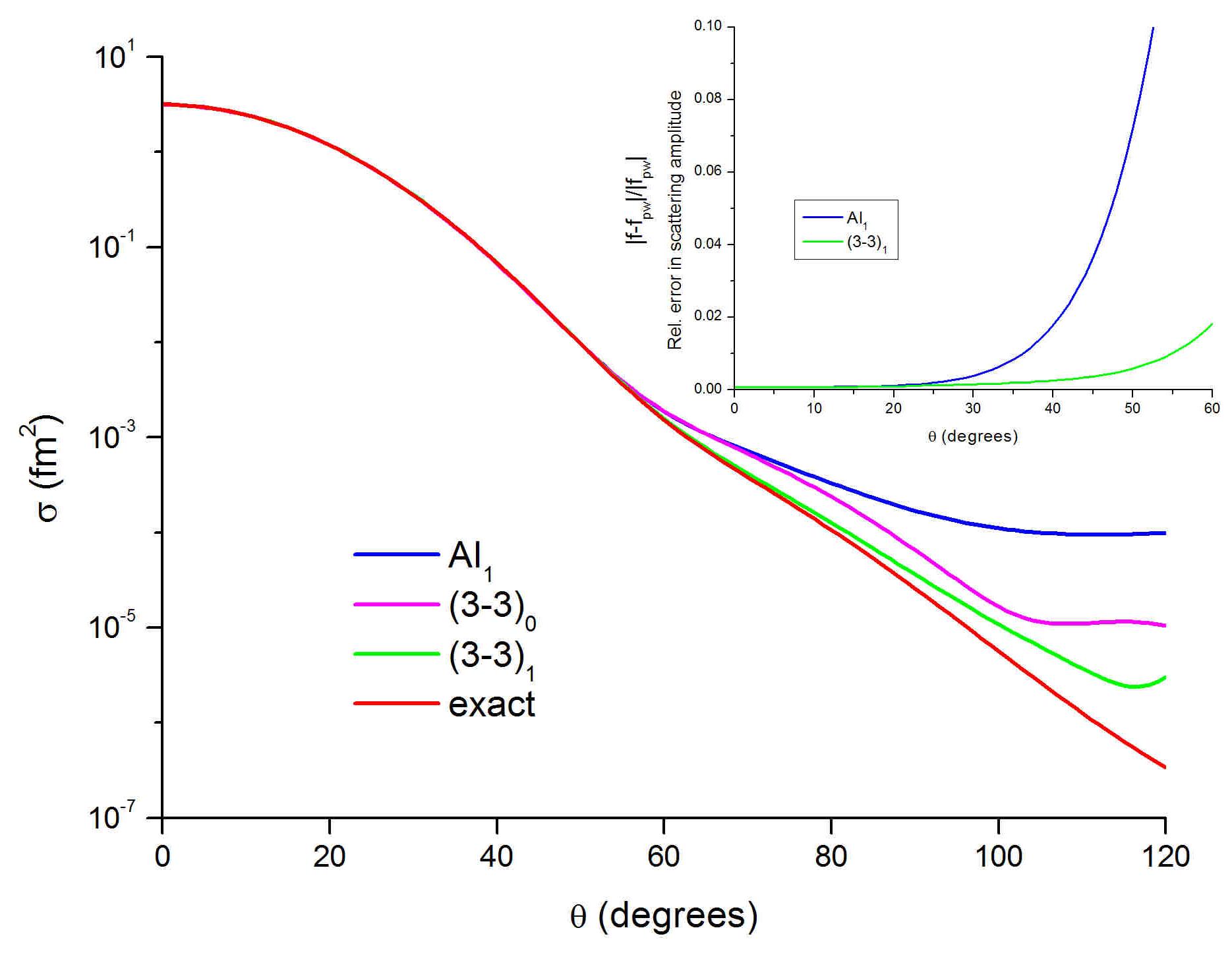}
		\caption {The differential cross section $\sigma$ as function of the scattering angle, for the Gaussian potential. The red line shows the exact, partial-wave results, the blue line the eikonal (AI) phase with the first correction. The other two lines show the variational approximation in the eikonal representation, without (magenta line) and with the first correction (green line). A sensible deviation is found at about 50 degrees, above which the variational approximation is then much closer to the exact result than the eikonal approximation. The upper panel shows the relative error in the scattering amplitude, with respect to the partial-wave calculations, for the eikonal approximation with first correction, and variational approximation with first correction (green line), for scattering angles up to sixty degrees. Note that at this value, the eikonal approximation is off by more than 10 percent, while the variational approximation still is within 2 percent.}
	\label{fig::sG1}
\end{figure}

\begin{figure}[htbp]
	\centering
		\includegraphics[width=10cm]{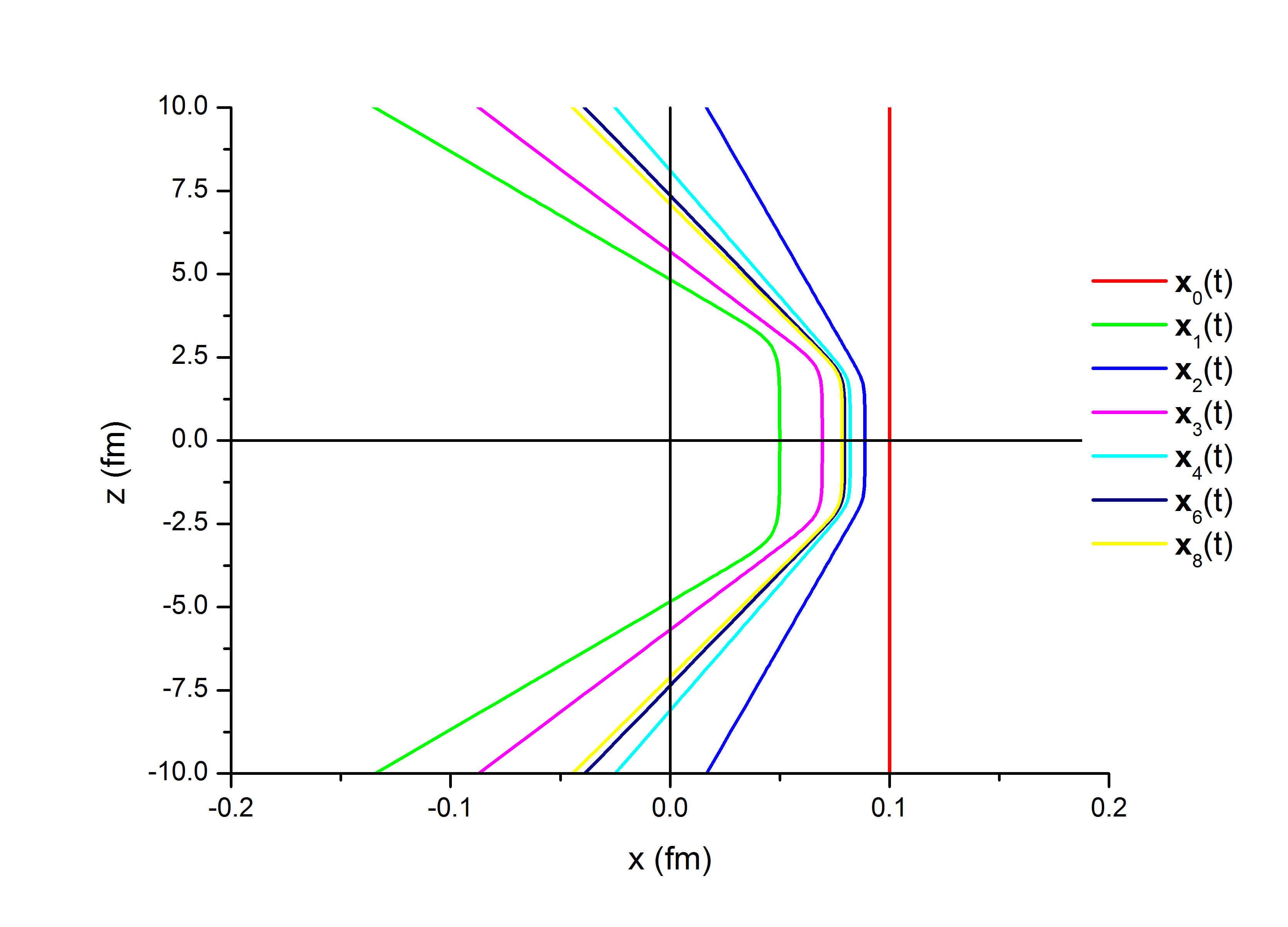}
		\caption {The same as figure \ref{fig::GTraj}, in the case of the Woods-Saxon potential. The significance of the parameters $R = 2$ fm (width of the potential well) and $a = 0.2$ fm (width of the edge of the well) can be clearly seen.}
	\label{fig::WTraj}
\end{figure}

This is done for the case, first, of a Gaussian potential of the form
\beq
V_G(x) = V_0 \exp\left(-\alpha x^2\right),\label{VG}
\enq
and second, for a Woods-Saxon potential
\beq
V_{WS}(x) = V_0 \left[ 1+\exp\lp \frac{x - R}{a}\rp\right]^{-1}.
\enq
The different parameters were chosen as
\beqa
k &=& 4\: \mathrm{fm}^{-1},\:m = 10\: \mathrm{fm}^{-1},\: V_0 = -0.2\: \mathrm{fm}^{-1}\\
\alpha &=& 1\: \mathrm{fm}^{-2},\:R = 2\: \mathrm {fm} \: \mathrm{and}\: a = 0.2\: \mathrm {fm}. 
\enqa
Using the relation $\hbar c \approx 200$ MeV fm$^{-1}$, these correspond to a particle of mass 2 GeV, with an initial kinetic energy of 160 MeV, interacting through a potential well of depth -40 MeV.\newline\noindent
We are in a high energy situation in both cases, and these parameters are precisely those for which the eikonal expansion was judged in \cite{Wallace} as being unsatisfactory. 
\newline \noindent For our calculation, the variational trajectory $\vecx(t)$ needs to be accurately found, for each scattering angle and $\vecb$. We determined it through iteration. Starting from the eikonal reference trajectory
\beq
\vecx_0(t) = \vecb + \frac \vecK m t,
\enq 
a sufficient number of iterations
\beq
\vecx_n (t) := \vecb + \frac \vecK m t -\frac{1}{2m}\int ds\:\nabla V(\vecx_{n-1}(s))|t-s|
\enq
was performed. Since the potentials are spherically symmetric, the motion takes place in both cases in the plane determined by $\vecb$ and $\vecK$. The x-axis was taken parallel to $\vecb$ and the z-axis parallel to $\vecK$. 
Figure \ref{fig::GTraj} shows the successive trajectories $\vecx_n(t)$ in the case of the Gaussian potential, for the given parameters and a particular choice of $b = 0.1$ fm. Note how the z-scale was stretched to see the effect on the asymptotic momenta. Figure \ref{fig::WTraj} shows the corresponding trajectories for the Woods-Saxon potential. To evaluate numerically the integrals needed in this iteration procedure, we used a Gauss-Legendre rule with 32 points and a sufficient number of subdivisions of the integration interval. The trajectories were numerically stored at points $t_i$, determined through the mapping
\beq
t_i = t_0\tan \psi_i, \quad \psi_i\in \left[-\frac \pi 2,\frac \pi 2\right],
\enq  
where the $\psi_i/(\pi/2)$ are the Gauss-Legendre points, and
\beq
t_0 = \frac m{k\sqrt\alpha}, \quad \mathrm{or}  \quad t_0 = \frac{m}{k}R,
\enq
is the typical time scale of the scattering process, for the Gaussian and Woods-Saxon potential respectively.
\paragraph{Gaussian Potential.}
With the help of the so-obtained trajectories, the phases $X_0$ and $X_1$ were evaluated according to equations (\ref{X033}) and (\ref{X133}), and the two-dimensional integration over the impact parameter $\vecb$ was reduced to a one-dimensional integration, using the fact that the phases depend only on $b$. All these integrations were performed using the Gauss-Legendre rule we mentioned above.\newline
The specific form of the Gaussian potential permits also the analytical computation of the first correction to the variational approximation without much trouble. We have already seen that the second cumulant takes the form
\beq
\lambda_2 = \la\chi_\vecK^2\rat -X_0^2 -2iX_1,
\enq
while in appendix \ref{2ndcumul} we evaluate, for the Gaussian potential,
\beqa
\la\chi_\vecK^2\rat =\nonumber&& V_0^2\dt1\dt2\left(\frac 1{1+\gamma_\parallel^2}\right)^{3/2}\exp\left( -\alpha\frac{\vecx^2(t_1)+\vecx^2(t_2)}{1+\gamma_\parallel^2}\right)\cdot\\
&&\cdot\exp\left( -2i\alpha\gamma_\parallel\frac{\vecx(t_1)\cdot\vecx(t_2)}{1+\gamma_\parallel^2}\right),\label{chi233}
\enqa
with 
\beq
\gamma_\parallel(|t_1-t_2|) = \frac \alpha m |t_1-t_2|. 
\enq
However, the oscillatory character of this second cumulant proved to be very troublesome in trying to evaluate it numerically. With the Gauss-Legendre rule we used for the evaluation of the first cumulant, it turned out not to be possible to achieve in reasonable time its computation to a sufficient accuracy. What we did to circumvent the problem was to use the numerical package DCUHRE \cite{dchure}, that uses an adaptive scheme for multi-dimensional integration, and permitted an accurate evaluation of the second cumulant for most values of $b$. For the largest values of $b$, we used a third method, that relies on the fact that the variational trajectory is essentially the same as the eikonal straight-line reference trajectory. More details can be found in appendix \ref{numlambda2}.\newline\newline\noindent
Results are shown in figure \ref{fig::sG1}. Presented in the plot is the exact cross section, evaluated with a sufficient number of partial waves, the eikonal (AI) approximation with the first correction included, and the variational approximation, with and without the first correction. The upper panel shows, on a linear scale, the relative deviation with respect to the exact scattering amplitude as a function of angle. Although the variational approximation, without the first correction, suffers the same deviation as the eikonal approximation at about fifty degrees, it clearly does better at larger angles. Especially, when the first correction is included, the improvement in this region over the eikonal approximation is impressive, as shown clearly by the upper panel. This is likely to be traced back to the appearance of the real term in $\lambda_2$.  

\paragraph{Woods-Saxon Potential.}
\begin{figure}[htbp]
	\centering
		\includegraphics[width=15cm]{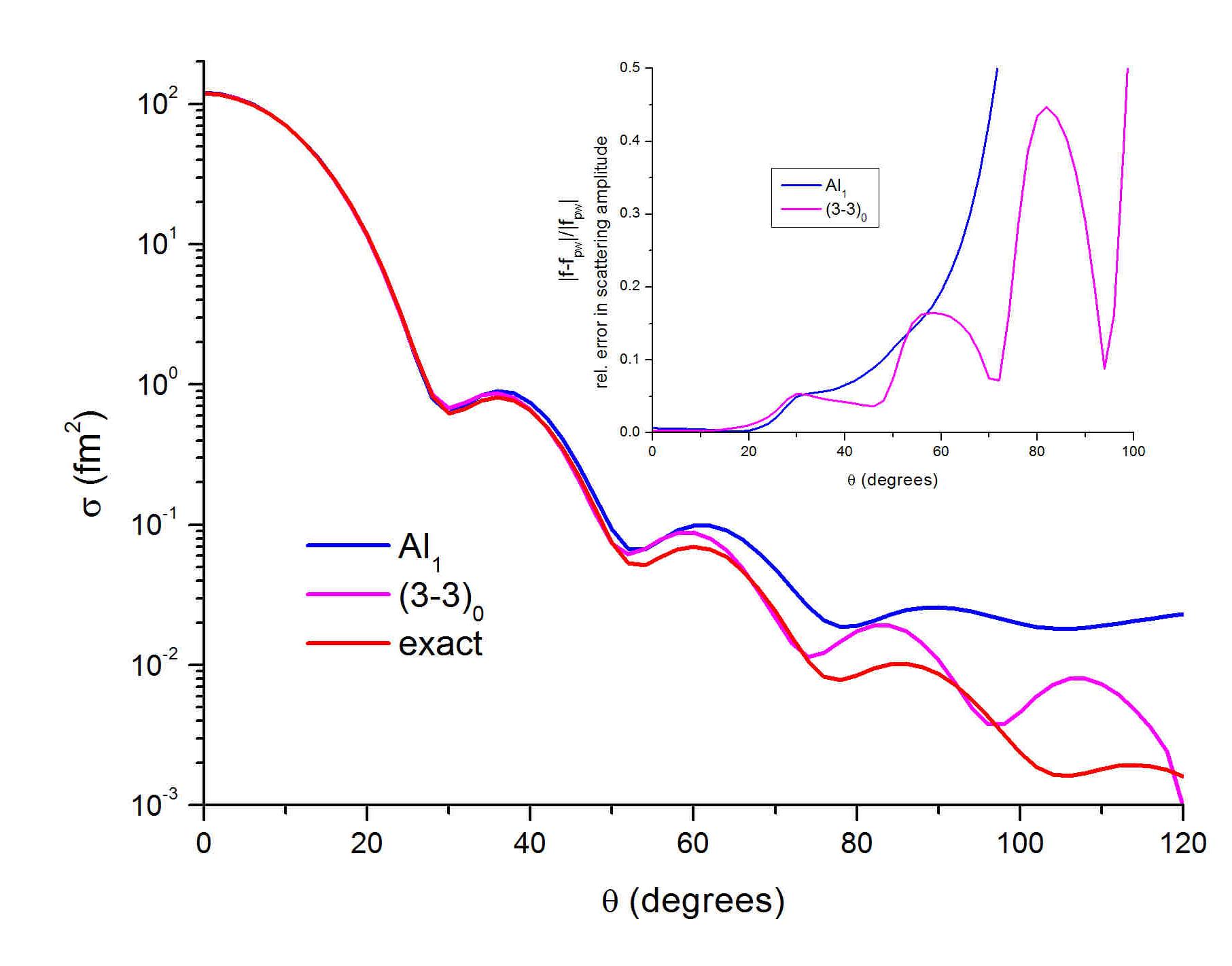}
		\caption {The differential cross section $\sigma$ as function of scattering angle in the case of the Woods-Saxon potential. Shown are the exact result given by the partial-wave expansion (red line), the eikonal (AI) phase with the first correction included, and the variational approximation in the eikonal representation (first cumulant only). Again, the variational approximation shows improvement at large angles. While the eikonal approximation is clearly inconsistent for such angles, the variational approximation is able to keep track of the diffractive structure, although it gets slowly out of phase. The upper panel shows the relative error in the scattering amplitude $f$ with respect to the partial-wave computations.}
	\label{fig::sWoods}
\end{figure}
We performed the same computations in the case of the Woods-Saxon potential, although we did not consider the first correction to the variational approximation. We used only Gauss-Legendre integration for each needed integral, with a sufficient number of points and subdivision of the integration intervals. Results are presented in figure \ref{fig::sWoods}.\newline\noindent Due to the edge in the Woods-Saxon potential, the cross section shows a diffraction structure, different from the Gaussian potential. While the eikonal approximation levels off and does not manage to reproduce the diffraction peaks at large angles, the variational approximation does it, although it gets more and more out of phase. The action of the first correction to the variational approximation would be interesting to study, albeit numerically quite involved.\newpage
\subsection{Ray Representation}
\begin{figure}[htbp]
	\centering
		\includegraphics[width=15cm]{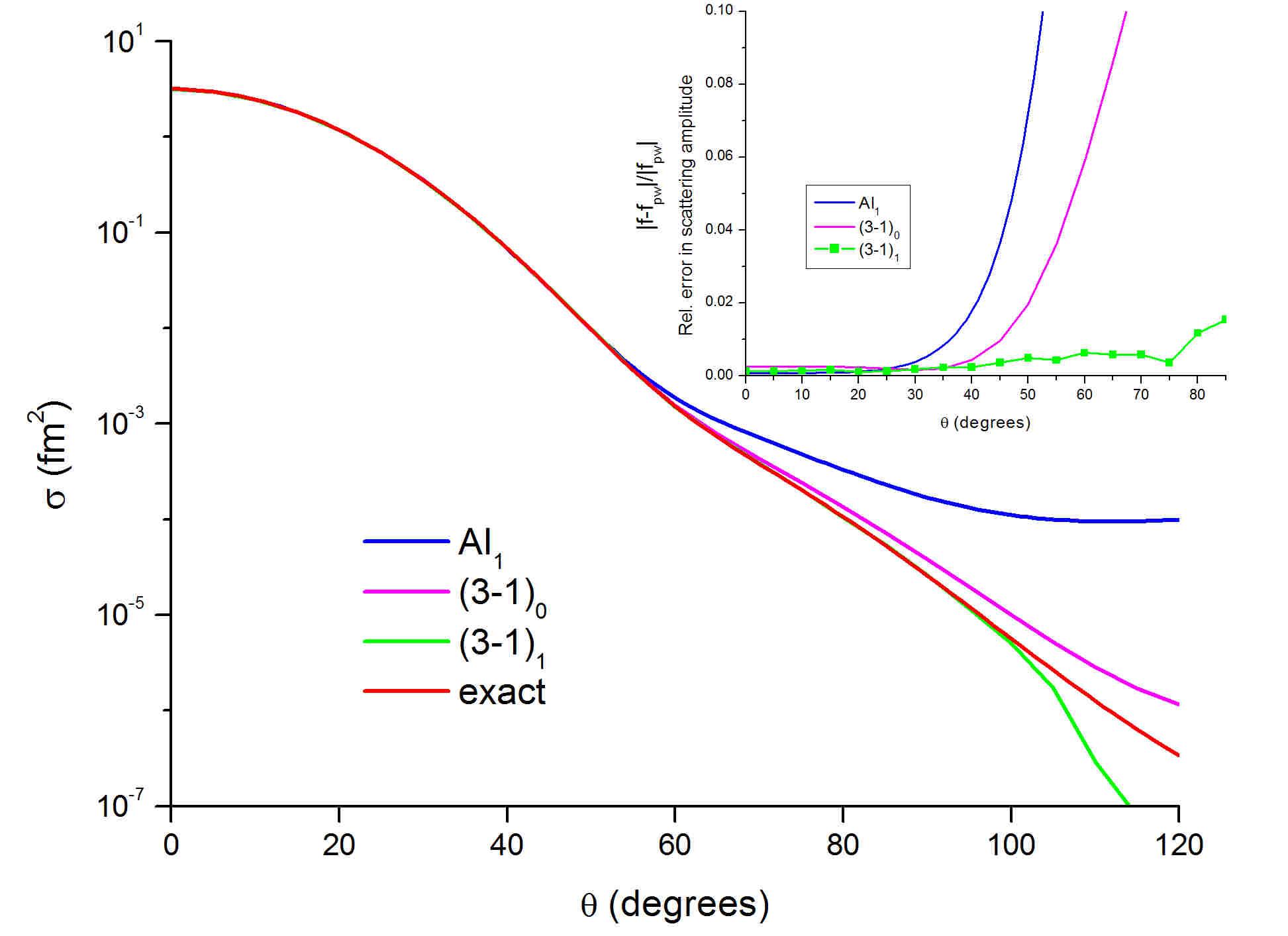}
		\caption {The differential cross section $\sigma$ as function of scattering angle in the case of the Gaussian potential. Shown are the exact result given by the partial-wave expansion (red line), the eikonal (AI) phase with the first correction included, and the variational approximation in the ray representation, without (magenta line) and with first correction (green line). The upper panel shows the relative deviation in the scattering amplitude. With the first correction included, the variational approximation in the ray representation is rather accurate, even at large angles.}
	\label{fig::sGray}
\end{figure}

We computed also the cross section in the ray representation in the case of the Gaussian potential. The variational trajectories were again obtained through iteration. Starting from
\beq
\quad \vecz_0(t) = \vecb + \frac \vecK m t + \frac \vecq {2m}|t|,
\enq
the successive approximations $\vecz_n(t)$ were found by setting
\beq
\vecz_n (t) := \vecb + \frac \vecK m t +\frac \vecq m |t| -\frac{1}{2m}\int ds\:\nabla \Vs s(\vecz_{n-1}(s))\left[|t-s| -|t|-|s|\right].
\enq
The same Gauss-Legendre rule as in the eikonal representation was used to find these trajectories. However, there are three significant differences with respect to the eikonal representation. First the motion is not in a plane anymore. Second, the trajectories are complex numbers, and third, one must perform numerically the full 2-dimensional integration over $\vecb$.\newline\noindent
Also, the effective potential $\Vs t$ is a new feature. For the Gaussian potential $V_G$, it is readily computed from equation (\ref{Veff}):
\beqa
\Vs t(\vecz(t)) &=& V_0\int \frac{d\vecp}{(2\pi)^3}\:\left(\frac{\pi}{\alpha}\right)^{3/2} \exp\left(-\frac 1 {4\alpha}\vecp^2 -\frac{1}{2}\sigma(t)\vecp^2_\perp + i \vecp\cdot\vecz(t)\right) \nonumber\\
&=& \frac {V_0}{1+2\alpha\sigma(t)}\exp\left(-\alpha\left[\vecz^2_\parallel +\frac {\vecz^2_\perp}{1 + 2\alpha\sigma(t)} \right] \right).
\enqa
Different from the case of the eikonal representation, we performed all the other integrations with the help of the DCUHRE package. The first correction is also obtained in appendix \ref{numlambda2}. \newline\noindent Results are plotted in figure \ref{fig::sGray}. The variational approximation in the ray representation without first correction (magenta line) shows, plausibly due to the presence of the real term in the scattering phases, a very similar behavior as its counterpart in the eikonal representation, but with the first correction included. We see our prediction that the ray representation will be more accurate clearly confirmed. With the first correction included, the result of the ray representation is even more impressive. As the upper panel shows, the approximation is within two percent of the exact result, all the way through ninety degrees, and only then begins to deviate.

%% file: Appendices.tex
\newpage
\appendix
\begin{center}                           
{\huge\bf Appendix}             
\end{center}   
\setcounter{equation}{0}
   
\section{Computation of $\chir n$}\label{chik}
In this section we calculate the expectation value $\chir n$ of the phase $\chi_\vecK$ in the ray representation, that was used without proof in the text:
\beqa\chir n &=&\nonumber(-1)^n\int\prodi i1n dt_i\int\prodi i1n \frac{d\vecp_i}{(2\pi)^3}\widetilde V(\vecp_i)\:\exp{\left(i\sumi i1n\vecp_i\left\{ \vecb + \frac{\vecp_{ray}(t_i)}{m}t_i\right\}\right)}\cdot\\
&\cdot&\exp\lp\frac{i}{4m}\sumi i1n \vecp_i\cdot\vecp_j \big[|t_i-t_j|-|t_i|_\perp-|t_j|_\perp\big]\rp.
\enqa
As in the text, whenever a $\perp$ symbol is found, like here as a companion of $|t_i|$ and $|t_j|$, it is meant that only the part that is perpendicular to $\vecK$ of the 3-dimensional vector that multiplies it is to be taken into account (i.e. $\vecp_i|t_i|_\perp = \vecp_{i\perp}|t_i|$). Similarly for the $_\parallel$ symbol.
We recall that in the ray representation, $\chi_\vecK$ is given by
\beq
\chi_\vecK(\vecb,\vecv,w) = -\int dt\:V\left(\vecb + \frac{\vecp_{ray}(t)}{m}t + \vecx_v(t) -\vecx_{v\perp}(t) - x_w(0)\right) .
\enq
The computation is very similar as in the case of a 3-dimensional antivelocity $\vecw$. We first start with a Fourier transformation of the potential, and solve the functional $\vecv$ and $w$ integrations. Let us consider first the $\vecv$ integration.
\beqa
&&\nonumber\int \mathcal D \vecv\: e^{i\frac m2 \lp\vecv,\vecv\rp+i\Bv}\exp\left( i\frac12\sumi i1n\vecp_i\int ds\:\vecv(s)\left[\sgn(t_i-s)-\sgn_\perp(s)\right]\right)\\
&=&\exp\left(-\frac i{2m}\int ds\:\left\{\vecB(s)+\frac12\sumi i1n \vecp_i\left[\sgn(t_i-s)-\sgn_\perp(-s)\right]\right\}^2\right).
\enqa
Similarly, the $w$ integration results in
\beq \exp\left(\frac i{2m}\int ds\:\left\{\C(s)-\frac12\sumi i1n \vecp_{i\parallel}\sgn(-s)\right\}^2\right).\enq
As already encountered, the $\vecB^2$ and $\C^2$ again cancel the normalisation $m_0$. Let us consider more closely the terms in the exponent that multiply $\vecp_i\cdot\vecp_j$ and are free of $\vecB's$ and $\C's$:
\beqa 
&&\int ds\:\nonumber\left(\sgn(t_i-s)\sgn(t_j-s) + 1_\perp -\sgn_\perp(-s)\left[\sgn(t_j-s)-\sgn(t_i-s)\right]-1_\parallel\right)\\
&=& -2\lp |t_i-t_j|-|t_i|_\perp-|t_j|_\perp\rp.
\enqa
Here we have used the relation (\ref{sgn}) only. It is thus seen that all divergent quantities cancel.
We further define the trajectory $\vecz(t)$ as
\beq
\vecz(t) = \vecb + \frac{\vecp_{ray}(t)}{m}t -\frac 1{2m}\int ds\:\vecB(s)\lp\sgn(t-s)-\sgn\perp(-s)\rp+\C(s)\sgn_\parallel(-s). 
\enq  
Getting the factors right, and plugging into $\chir n$, it reduces to
\beqa
\chir n &=&\nonumber(-1)^n\int\prodi i1n dt_i\int\prodi i1n \frac{d\vecp_i}{(2\pi)^3}\widetilde V(\vecp_i)\exp\left(i\sumi i1n \vecp_i\cdot\vecz(t_i)\right)\cdot\\
&\cdot&\exp\lp\frac{i}{4m}\sumi i1n \vecp_i\cdot\vecp_j \big[|t_i-t_j|-|t_i|_\perp-|t_j|_\perp\big]\rp,
\enqa
which was the formula to be proven.

\setcounter{equation}{0}
\section{Reduction Formulae}
\subsection{Reduction of $\chi_1(b)$}\label{reductiontau1}

We prove in this section the formula for the first order phase $\chi_1(b)$, for spherically symmetric potentials, which was mentioned in the text:
\beq \chi_1(b) = -\frac 1{v^2K} \lp 1 + b\frac{d}{db}\rp\int_0^{\infty}dZ\: V^2(r).\enq
Since we are dealing with a spherically symmetric potential, we will always write, in such expressions, $V(r_i)$ instead of $V(\sqrt{b^2+Z_i^2})$, for the sake of clarity.
To this aim, we start from the general formula (\ref{tau1}), and effectuate the substitutions $Z_1 = \frac K m t$, and $Z_2 = \frac K m s$. We then obtain 
\beq\chi_1(b) = \frac  1{4Kv^2} \int dZ_1\:\int dZ_2\:\nabla V(r_1)\cdot\nabla V(r_2)|t-s|.\enq
Since $\vecb$ and $\vecK$ are perpendicular to each other, and $\vecK$ points along the third axis, it is immediately
recognized that $\pa 3 V(\vecb + Z\hat\vecK) = \pa Z V(r)$. We furthermore choose $\vecb$ to point along the first axis, so that $\pa 1 V(r) = \pa b V(r)$ also holds.
We will make extensive use in the following calculations of the relation $Z\pa b V(r) = b \pa Z V(r)$, which is a special case of $x_i\pa j = x_j\pa i$, that holds for any spherically symmetric function.\newline
We thus have
\beq
\chi_1(b) = \frac  1{4Kv^2} \int dZ_1\:\int dZ_2\:\left(\pa {Z_1} V(r_1)\pa {Z_2}V(r_2) + \pa b V(r_1)\pa b V(r_2)\right)|Z_1-Z_2|.
\enq
We are allowed to integrate over $Z_1$ and $Z_2$ by parts in the first term, while for the second term we first write $(Z_1-Z_2)\:\sgn(Z_1-Z_2)$ for $|Z_1-Z_2|$, and use $Z\pa b V(r) = b \pa Z V(r)$.
We obtain
\beq\chi_1(b) = \frac  1{4Kv^2} \int dZ_1\: \left\{-2V^2(r) +2b\int dZ_2\: \pa {Z_1} V(r_1)\pa bV(r_2)\sgn(Z_1-Z_2)\right\},\enq
 where we have used
\beq
\pa {Z_1}\pa{Z_2}|Z_1-Z_2| = -\pa {Z_1}\sgn(Z_1-Z_2) = -2\delta(Z_1-Z_2).
\enq
The second term can now also be integrated by parts, and, after recognizing that
\beq 2V(r)\pa b V(r) = \pa b V^2(r),\enq
one finally has
\beq \chi_1(b) = \frac  1{4Kv^2} \int dZ\: \left(-2V^2(r) - 2b\frac{d}{db}V^2(r)\right).\enq
If we use the symmetry of the integrand to integrate from $0$ to $\infty$, one gets exactly what was to be proven:
\beq \chi_1(b) =- \frac 1{v^2K} \lp 1 + b\frac{d}{db}\rp\int_0^{\infty}dZ\: V^2(r).\enq

\subsection{Reduction of $\omtwovar(b) $} \label{omega2}
In this section we derive the reduced expression of $\omtwovar(b)$, for spherically symmetric potentials, that was mentioned in the text :
\beq \nonumber \omtwovar(b) = \frac{1}{8K^2}\left[2\chi_0^{'2}+\chi_0^{''}\left(\chi_0+b\chi_0^{'}\right)+\frac 1 b \chi_0 \chi_0^{'}\right].
\enq
This $\omtwovar$ is given by the first term in the expansion of the second cumulant, evaluated along the straight line trajectory,
\beq \omtwovar(b) = - \frac{1}{16v^2K^2}\dZ1\dZ2(Z_1-Z_2)^2\pa i\pa j V(\vecb+Z_1\hat{\vecK})\pa i\pa j V(\vecb+Z_2\hat{\vecK}),
\enq
where a sum over the $i$ and $j$ indices is implicit. \newline
We will use the same conventions as in appendix \ref {reductiontau1}. 
The integral consists of a few terms that need to be calculated. Let us consider them one by one.
 
\paragraph{case $i=j=2$:} We will call this term $A_1$. Since
\beq \pa i \pa j V(r) = V^{''}(r)\frac{x_ix_j}{r^2} + V'(r)(\frac{\delta_{ij}}{r} - \frac{x_ix_j}{r^3}),\enq  it is the only non-zero term where one of the two indices takes the value two. It reads then
\beqa A_1(b) &=& \dZ1\dZ2(Z_1 - Z_2)^2\:\frac{V'(r_1)}{r_1}\frac{V'(r_2)}{r_2}\\
&=&2\dZ1 Z_1^2 \frac{V'(r_1)}{r_1}\dZ2\frac{V'(r_2)}{r_2}.\enqa
We used in the second line the symmetry properties of the integrand.
The $Z_2$ integral is, up to a factor of $b$ and $v$, the derivative of $\chi_0(b)$, and the first can be integrated by parts with the help of $Z\:V'/r = \pa Z V$, so that we obtain
\beq A_1(b) = -2\frac{v^2}{b}\chi_0'\chi_0.\enq

\paragraph{case $i=j=b$:} This term, that we call $A_2$, reads
\beqa A_2 &=& \dZ1\dZ2(Z_1-Z_2)^2\pa b \pa b V(r_1) \pa b \pa b V(r_2)\\
&=& 2\dZ1 Z_1^2\pa b \pa b V(r_1)\dZ2 \pa b \pa b V(r_2).\enqa
The second integral is recognized to be the second derivative of $\chi_0$, while the first, using $Z \pa b\pa b V = (1 + b\pa b) \pa ZV$, is integrated by parts. We get
\beqa A_2(b) &=& v\chi_0^{''}\int dZ\:\left[1+b\pa b\right]V(r)\\
&=&-2v^2\chi_0^{''}(\chi_0+b\chi_0').\enqa

\paragraph{case $(i,j) =(b,Z)\:\: \mathrm{and}\:\: (Z,b)$:} These two terms are again readily computed with integration by parts :
\beqa A_3(b) &=& 2\dZ1\dZ2 (Z_1-Z_2)^2 \pa b \pa b V(r_1) \pa b \pa b V(r_2)\\
&=&-4\left(\dx Z Z\pa Z\pa b V(r) \right)^2\\
&=&-4v^2\chi_0^{'2}.\enqa
Since any other term than those three, obviously vanishes, we readily have proven our claim :
\beqa \nonumber \omtwovar(b) &=& A_1(b) + A_2(b) + A_3(b) \\
&=& \frac{1}{8K^2}\left[2\chi_0^{'2}+\chi_0^{''}\left(\chi_0+b\chi_0^{'}\right)+\frac 1 b \chi_0 \chi_0^{'}\right].
\enqa
\subsection{Reduction of $\tautwo(b) $} \label{tau2}

This section is devoted to the derivation of the representation of $\tautwo$, in the case of a spherically symmetric potential function, that was presented in the text:
\beqa \tautwo(b) = \nonumber &-&\frac{1}{v^3K^2}\left(1 +\frac53b\frac{d}{db}+\frac13b^2\frac{d^2}{db^2}\right)\int_0^{\infty}dZ\:V^3(r) - \\
&-&\frac{1}{8K^2}\left[\chi_0(\chi_0^{'})^2+\frac13b(\chi_0^{'})^3\right].\enqa
We will use throughout this section the same notation convention as in the last section.
The expression to be reduced is
\beq \tautwo = -\frac1{8v^3K^2}\dZZZ \pa i \pa j \V1 \pa i \V2\pa j \V3\dabs12\dabs13.\enq
It is seen immediately that no terms with an index $i$ or $j = 2$ contributes to the sum. Thus, there are three terms left to be computed.
\paragraph{case $i = j = Z$:} This term, that we define as $A_1(b)$, is the less involved. We integrate by parts to obtain :
\beqa A_1(b) &:=& \nonumber \dZZZ \pa {Z_1}\pa {Z_1}\V1 \pa {Z_2}\V2 \pa {Z_3} \V3\dabs12\dabs13\\
&=&\nonumber  \dZZZ \pa {Z_1}\pa {Z_1}\V1 \V2 \V3\dsgn12\dsgn13\\
&=&\nonumber -2 \dZZ \pa {Z_1}\V1 \V2 \V1\dsgn12
\enqa
There we only made extensive use of the rule $|x|' = \sgn(x)$ and $\sgn(x)' = 2\delta(x)$. A last integration by parts permits us to conclude
\beq A_1(b) =  4\int dZ\: V^3(r).\enq

\paragraph{case $(i,j) = (b,Z)\:\mathrm{and} \:(Z,b)$:} These two terms consist of
\beq A_2(b) := 2\dZZZ \pa {Z_1}\pa b \V1 \pa {Z_2} \V2 \pa{b} \V3 \dabs12\dabs13.\enq 
There we will encounter a function that plays an important role. We define it by the following equation,
\beq 
X(Z,b) := \int dZ_1\:\sgn(Z-Z_1)V(r_1).
\enq
It has two nice properties, which we will use many times: 
\beq X(\pm\infty,b) = \mp v\chi_0(b)\quad\mathrm{and}\quad\pa Z X(Z,b) = 2V(r(Z,b)).\enq
From now on we will not write explicitly the b-dependence of X, and write a dot for $\pa b$.\newline
After one integration by parts over $Z_2$, one obtains
\beqa A_2(b) &=& \nonumber2\int dZ_{1,3}\: X(Z_1)\V1 \dot V(r_3)\pa {Z_1}\dot V(r_1)\dabs13\\
&=&-4\int dZ\nonumber \:X(Z)\dot X(Z)\dot V(r)-2\int dZ_{1,3}\:\dot V(r_1)\dot V(r_3)\V1\dabs13.
\enqa
We leave the term on the left as it is for now. The second term can be further reduced. To this aim, we decompose $\dabs13$ in $\dsgn13(Z_1-Z_3)$, and we effectuate the integration using $Z\pa b = b \pa Z$. After some algebra, one obtains finally
\beq A_2(b) = \frac{12}{3}b\frac{d}{db}\int dZ\:V^3(r) -2F(b),\enq
with
\beq F(b) := -\int dZ\:X(Z)\dot X(Z)\dot V(r).\enq
\paragraph{case $i = j = b$:}This quantity reads
\beq
A_3(b) := \dZZZ \dabs13\dabs12\ddot V(r_1)\dot V(r_2)\dot V (r_3).\enq
First we consider the $Z_1$ and $Z_3$ integrations. We have
\beqa \int dZ_2\:\dot V(r_2)\dabs12 &=& \int dZ_2\:\dot V(r_2)\dsgn12(Z_1 -Z_2)\\
&=&Z_1\dot X(Z_1) -2b\V1,\enqa so that $A_3$ can be rewritten as
\beqa A_3(b) &=& \int dZ\:\ddot V(r)\left(Z\dot X (Z)-2bV(r)\right)^2 
\enqa Further algebra involving integration by parts then reduces $A_3$ to
\beq A_3(b) = \frac43b\frac{d}{db}\int dZ\:V^3(r) +\frac43b^2\frac{d^2}{db^2}\int dZ\:V^3(r) + 4G(b) + H(b),\enq where $G$ and $H$ are defined as
\beqa G(b)&=&b^2\int dZ\:\dot V(r)\dot X(r)\pa Z V(r)\\
H(b) &=& \int dZ\:Z(1+b\pa b)(\pa Z V(r))\dot X^2(Z).\enqa
\newline
The last step is to recognize that $4G-2F+H$ is none other than
\beq 4G-2F+H = \chi_0(\chi_0^{'})^2+b(\chi_0^{'})^3+\frac43\frac{d}{db}\int dZ\: V^3(r).\enq
This is done of course also by integrating by parts, but this time the integrand evaluated at infinity contributes, making $\chi_0$ and its derivative to appear. By summing over the different quantities so computed, one obtains exactly what was claimed.
\setcounter{equation}{0}
\section{The Second Cumulant for a Gaussian Potential.} \label{2ndcumul}
In this section the computation of $\la\chi^2_k\rat$ is done in the case of a Gaussian potential of the form
\beq V(\vecx) = V_0\exp\lp-\alpha\:\vecx^2\rp.\enq
The second cumulant is then, in both the eikonal and ray representations, given by subtracting the phases $X_0^2$ and $2iX_1$ in their respective representations (see equations (\ref{l233}) and (\ref{l231})).
After plugging this form of the potential into the formulas for $\la\chi^2_k\rat$ we already derived, we get
\beq 
\la\chi_\vecK^2\rat = V_0^2\left(\frac{\pi}{\alpha}\right)^3\int dt_1\int dt_2\int \dpun\int\dpdeux\:\chi_\vecK(t_1,t_2,\vecp_1,\vecp_2)
\enq
in both cases, with
\beqa
&&\chi_\vecK^{(3-3)}(t_1,t_2,\vecp_1,\vecp_2)\nonumber\\
&=& \exp \left\{ -\frac{1}{4\alpha}\left(\vecp_1^2+\vecp_2^2\right)+ i\vecp_1\cdot\vecz(t_1)+i\vecp_2\cdot\vecz(t_2) + \frac i{2m}\vecp_1\cdot\vecp_2|t_1-t_2|\right\}\nonumber,
\enqa
and 
\beqa
&&\chi_\vecK^{(3-1)}(t_1,t_2,\vecp_1,\vecp_2)\nonumber = \chi_\vecK^{(3-3)}(t_1,t_2,\vecp_1,\vecp_2)\cdot\\
&\cdot& \exp\left\{ -\frac i{2m}\left(\vecp_{1\perp}^2|t_1|+\vecp_{2\perp}^2|t_2|\right)-\frac i{2m}\vecp_{1\perp}\cdot\vecp_{2\perp}\left(|t_1|+|t_2|\right)\right\}.\nonumber
\enqa
In these expressions, $\vecz(t)$ means the reference trajectory that we denoted $\vecx(t)$ in the eikonal representation, or by the same letter $\vecz$ in the (3-1) case. These are 6-dimensional Gaussian integrals and can be thus given in closed form. In order to effectuate this more efficiently in both cases, we first consider the integral over one component of $\vecp_1$ and the corresponding component of $\vecp_2$. These are all integrals of the form
\beq \int \frac {dp_1}{2\pi}\:\int \frac{dp_2}{2\pi}\:\exp\left(-a_1p_1^2-a_2p_2^2 + ip_1z_1+ip_2z_2+bp_1p_2\right),\enq
with different coefficients $a$ and $b$ for the perpendicular and parallel components. These 2-dimensional integrals are easily done and are equal to
\beq \label{2dint}\frac1{2\pi}\exp\lp -\frac{a_2z_1^2+a_1z_2^2+bz_1z_2}{4a_1a_2-b^2}\rp\sqrt{\frac 1{4a_1a_2-b^2}}.\enq 
\paragraph{Eikonal Representation:} 
The eikonal representation does not distinguish between parallel and perpendicular components, so that for each pair of integrals,
\beq a_1 = a_2 = -\frac1{4\alpha}\quad \mathrm{and}\quad b = \frac i{2m}|t_1-t_2|.\enq
Besides, we define a new quantity $\gamma_\parallel$ as being
\beq \gamma_\parallel(t_1,t_2) := \frac\alpha m|t_1-t_2|,\enq
such that $4a_1a_2 -b^2$ is then simply $(1+\gamma_\parallel^2)/4\alpha^2$. Equation (\ref{2dint}), applied to each component of $\vecp_1$ and $\vecp_2$, then says
\beqa
\la\chi_\vecK^2\rat =\nonumber&& V_0^2\dt1\dt2\left(\frac 1{1+\gamma_\parallel^2}\right)^{3/2}\exp\left( -\alpha\frac{\vecz^2(t_1)+\vecz^2(t_2)}{1+\gamma_\parallel^2}\right)\cdot\\
&&\cdot\exp\left( -2i\alpha\gamma_\parallel\frac{\vecz(t_1)\cdot\vecz(t_2)}{1+\gamma_\parallel^2}\right).
\enqa
\paragraph{Ray Representation:} 
In this case, different coefficients $a$ and $b$ correspond to the parallel and perpendicular components. The integration over the parallel components is the same as in the eikonal representation, but for the perpendicular components, we have
\beqa
a_1 = \frac 1 {4\alpha}+\frac{i}{2m}|t_1|,\quad a_2 = \frac 1{4\alpha}+\frac{i}{2m}|t_2|,\quad b = \frac{i}{2m}\lp|t_1-t_2|-|t_1|-|t_2|\rp.
\enqa
With the introduction of
\beq \gamma_{1,2}(t_{1,2}) := \frac\alpha m|t_{1,2}|,\quad\mathrm{and}\quad\gamma_3(t_1,t_2) := \gamma_\parallel-\gamma_1-\gamma_2,\enq we have
\beq 4a_1a_2 -b^2 = \frac{1}{4\alpha^2}\left[\lp 1+2i\gamma_1\rp\lp 1+2i\gamma_2\rp+\gamma_3^2\right],\enq
and finally, according to equation (\ref{2dint}),
\beqa
&&\la\chi_\vecK^2\rat_{ray} =V_0^2\dt1\dt2\frac {1}{\sqrt{1+\gamma_\parallel^2}}\left[\lp 1+2i\gamma_1\rp\lp 1+2i\gamma_2\rp+\gamma_3^2\right]^{-1}\nonumber \\
\nonumber &&\cdot\exp\left(-\alpha\frac{\vecz^2_{\perp}(t_1)(1+2i\gamma_2)+\vecz^2_{\perp}(t_2)(1+2i\gamma_1)}{\lp 1+2i\gamma_1\rp\lp 1+2i\gamma_2\rp+\gamma_3^2}     \right)\\
&&\cdot\nonumber\exp\left(-2i\alpha\gamma_3\frac{\vecz_{\perp}(t_1)\cdot\vecz_{\perp}(t_2)}{\lp 1+2i\gamma_1\rp\lp 1+2i\gamma_2\rp+\gamma_3^2}     \right)\\
 &&\cdot\exp\left(-\alpha\frac{\vecz^2_\parallel(t_1)+\vecz^2_\parallel(t_2)}{1+\gamma_\parallel^2}     \right)\exp\lp   -2i\alpha\gamma_\parallel\frac{\vecz_\parallel(t_1)\cdot\vecz_\parallel(t_2)}{1+\gamma_\parallel^2}\rp.\enqa
\setcounter{equation}{0}
\section{Angle Corrections to the 2nd Order Scattering Phases}
\label{2order}

The aim of this section is to derive two results that were mentioned in the text that contributed to the second order scattering phases
$\omega_2$ and $\tau_2$ in a high-$k$ expansion of the scattering matrix in the eikonal representation. The phases $\tauk(b)$ and $\omk(b)$ were defined as the imaginary and real part of the second order term arising because of the presence of the $\cos(\theta/2)$ term in the expression (\ref{T33}) for the T-matrix in the eikonal representation. Similarly, the imaginary and real part of the second order term arising because of the presence of the very same term in the leading phase were named $\taukk(b)$ and $\omkk(b)$. The identities proven in this section are
\beq \label{un}
i\tauk -\omk= \frac{1}{8k^2}\lp i\Delta_b\chi_0 - \left(  \nabla_b\chi_0\right)^2\rp
\enq
and
\beq \label{deux}
i\taukk -\omkk= \frac{1}{8k^2}\lp 2\:\left(\nabla_b\chi_0\right)^2+\chi_0\Delta_b\chi_0 +i ((\nabla_b\chi_0)^2\chi_0 - \Delta_b\chi_0) \rp,\enq
where $\chi_0$ denotes the standard eikonal phase
\beq \label{trois}
\chi_0(b) = -\frac mk \int dZ\: V(r(b,Z)).
\enq 
We first prove (\ref{un}).
To this aim, we first recognize that, up to second order in $1/k$, $\cos(\theta/2) = (1-q^2/8k^2)$. This permits us to rewrite the T-matrix as (in the subsequent development an equal sign always means equality up to terms of order four and higher.)
\beq
T_{i\rightarrow f}^{(3-3)} = i\frac{k}{m} \int d^2b\:\lp1-\frac{q^2}{8k^2}\rp e^{-i\vecq\cdot\vecb}\lp e^{i\chi_0}-1\rp,
\enq
Since we are interested in the second order terms, it is sufficient to consider only the leading phase $\chi_0$ in the exponential function.
The factor $q^2$ can be trivially converted to a Laplacian operator, and then brought on the other side by an integration by parts.
\beq
T_{i\rightarrow f}^{(3-3)} = i\frac{k}{m} \int d^2b\: e^{-i\vecq\cdot\vecb}\lp1+\frac{\Delta_b}{8k^2}\rp\lp  e^{i\chi_0}-1\rp.
\enq 
Since the identity
\beq \Delta_be^{i\chi_0} = \nabla_b\lp e^{i\chi_0}i\nabla_b\chi_0\rp = e^{i\chi_0}\lp i \Delta_b\chi_0-\lp\nabla_b\chi_0\rp^2\rp,
\enq
holds, one obtains
\beqa
T_{i\rightarrow f}^{(3-3)} &=& i\frac{k}{m} \int d^2b\: e^{-i\vecq\cdot\vecb}\lp1+\frac{1}{8k^2}\left[ i \Delta_b\chi_0-\left(  \nabla_b\chi_0\right)^2\right]\rp e^{i\chi_0}-1.\\
&=&\nonumber i\frac{k}{m} \int d^2b\: e^{-i\vecq\cdot\vecb}\lp \:\exp\left\{i\chi_0 + \frac{1}{8k^2}\left[ i \Delta_b\chi_0-\left(  \nabla_b\chi_0\right)^2\right]\right\} -1 \rp,
\enqa 
which is the first of the expected results.
To prove (\ref{deux}), the procedure is very similar. From (\ref{trois}) it follows that
\beq 
\chi_0 = \chi_0\left(1+\frac {q^2} {8k^2}\right),
\enq
which in turn implies
\beqa
T_{i\rightarrow f}^{(3-3)} &=& i\frac{K}{m} \int d^2b\: e^{-i\vecq\cdot\vecb}\lp \exp\left\{i\chi_0\lp1+\frac{q^2}{8k^2}\rp\right\}-1\rp\\
&=&i\frac{K}{m} \int d^2b\:e^{-i\vecq\cdot\vecb}\lp e^{i\chi_0}\lp1+i\chi_0\frac{q^2}{8k^2}\rp-1\rp.
\enqa 
Again, equalities are to be understood to hold up to fourth order. We now use the same trick as we just did, i.e. we convert the $q^2$-term to a Laplacian operator and integrate by parts:
\beq
T_{i\rightarrow f}^{(3-3)} = i\frac{K}{m} \int d^2b\: e^{-i\vecq\cdot\vecb}\left\{ e^{i\chi_0}\left(1 - i\frac{\Delta_b}{8k^2}\right)\left[\chi_0\: e^{i\chi_0}\right]-1\right\}.
\enq 
A small calculation shows
\beq
\Delta_b\left[ \chi_0\: e^{i\chi_0}\right] = e^{i\chi_0}\left(-\left(  \nabla_b\chi_0\right)^2\chi_0+\Delta_b\chi_0+i\left(2\left(  \nabla_b\chi_0\right)^2+\chi_0\Delta_b\chi_0\right)\right),\enq and subsequently,

\beq
T_{i\rightarrow f}^{(3-3)} = i\frac{K}{m} \int d^2b\: e^{-i\vecq\cdot\vecb}\lp e^{\left(i\chi_0- \frac{i}{8k^2}\left[-\left(  \nabla_b\chi_0\right)^2\chi_0+\Delta_b\chi_0+i\left(2\left(  \nabla_b\chi_0\right)^2+\chi_0\Delta_b\chi_0\right)\right]\right)}-1\rp,
\enq 
which is what was to be proven.
\setcounter{equation}{0}
\section{Numerical Evaluation of the Second Cumulant} \label{numlambda2}
{In this section, we briefly explain the difficulties encountered in evaluating numerically the second cumulant for the variational approximation in the ray representation for the Gaussian potential (\ref{VG}). As we already said in the section on numerical results, we were unable to evaluate the main component $\la\chi_\vecK^2\rat$ of the second cumulant accurately, through Gauss-Legendre integration. Since the Gauss-Legendre points enter  our program in every integration that had to be computed, the requested massive increase of the number of points would be too costly in time to be performed. Instead, we chose to adopt another integration scheme, the adaptive integration package DCUHRE, to calculate the second cumulant. Since the variational trajectories were stored only at places corresponding to the Gauss-Legendre points, a linear interpolation between these points was also done. As a typical example of the problem encountered, figure \ref{fig::NumCumul} shows the modulus of the real part of $\la\chi_\vecK^2\rat$ computed in these two different ways (blue and red curves), for a special value of $K = 2\:\mathrm{fm}^{-1}$, and for integration parameters adapted to our situation, as a function of $b$. It shows that two orders of magnitude in precision were gained through this procedure.\newline\noindent However, this was still not enough, since the adaptive integration also levels off after a given value of $b$, that was still to be accounted for in the computation of the differential cross section. To solve this problem we used the fact the trajectory is sufficiently close to a straight line to approximate $\la\chi_\vecK^2\rat$ by its value given by (\ref{chi233}), with
\beq \vecb +\frac \vecK m t \quad\ \text{instead of}\quad \vecx(t).
\enq
In that case, after passing to the variables
\beq 
T := \frac 12 (t_1 + t_2),\quad\mathrm{and}\quad t := t_1-t_2,
\enq      
the integral over $T$ is a pure Gaussian and can be analytically solved. The 1-dimensional integration that is left was then numerically evaluated and compared to the full 2-dimensional integration involving the variational trajectory. It leads to the magenta line in figure \ref{fig::NumCumul}. As one can see, the agreement between this curve and the one representing the two-dimensional adaptive integration is good enough, before the adaptive integration fails. 
\begin{figure}[htbp]
	\centering
		\includegraphics[width=15cm]{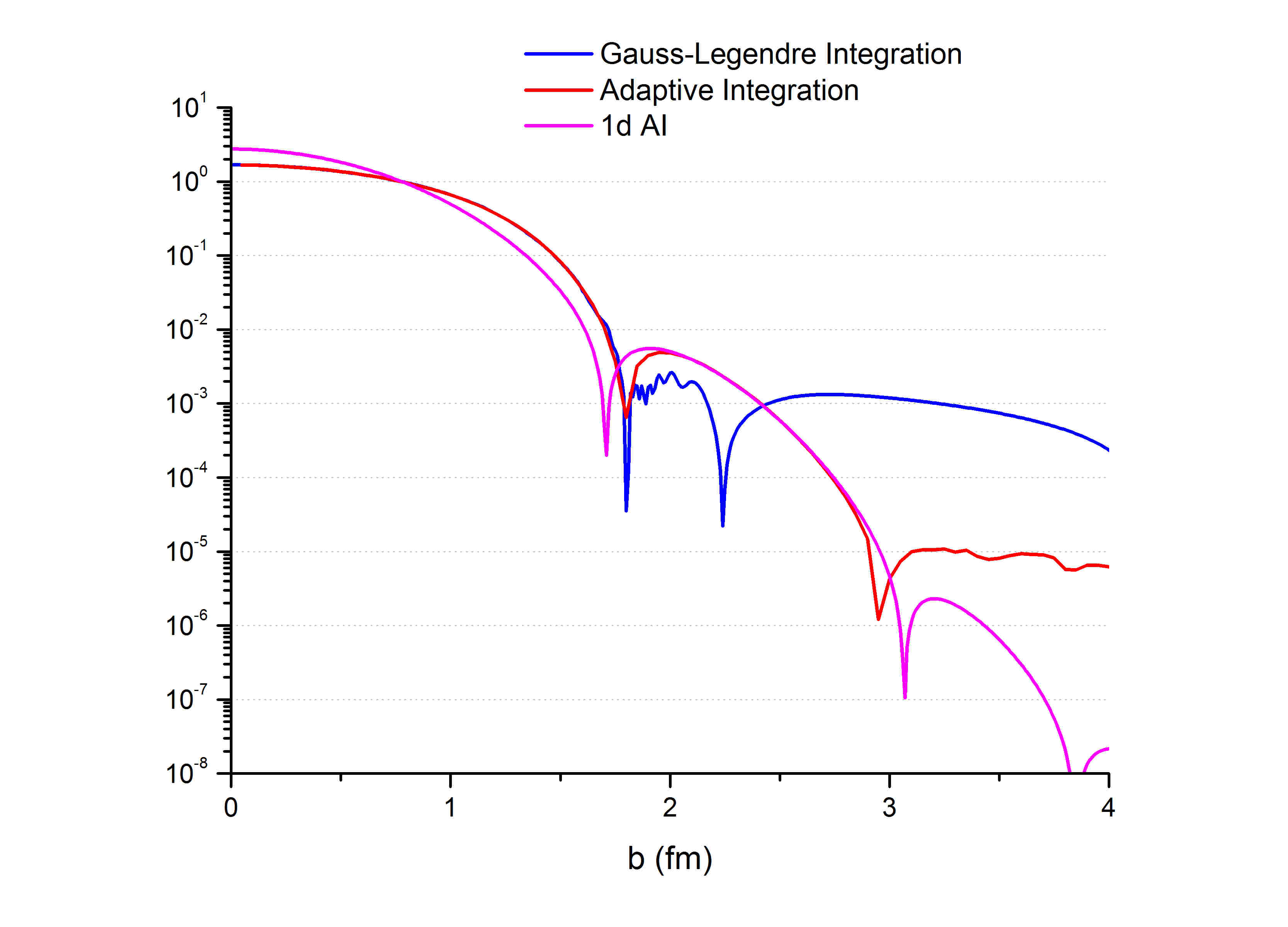}
		\caption {The modulus of the real part of $\la\chi_\vecK^2\rat$ for the Gaussian potential, as a function of b, evaluated numerically by Gauss-Legendre integration and adaptive integration, with optimal control and accuracy parameters. The third, magenta line, is the 1-dimensional version, as explained in the text. Both the Gauss-Legendre as well as the adaptive integration proved unable to perform the integration up to sufficiently large values of $b$, although the adaptive integration levels off two orders of magnitudes later.}
	\label{fig::NumCumul}
\end{figure}
}
